\documentclass[prl,amsmath,amssymb,twocolumn,showpacs,10pt]{revtex4-2}
\usepackage{amsfonts}
\usepackage{amsthm}
\usepackage{color}
\usepackage{amsmath}
\usepackage{graphicx}
\usepackage{amssymb}

\begin{document}

\newcommand \be  {\begin{equation}}
\newcommand \bea {\begin{eqnarray} \nonumber }
\newcommand \ee  {\end{equation}}
\newcommand \eea {\end{eqnarray}}

\title{ Eigenfunction non-orthogonality factors and the shape of CPA-like dips in a single-channel reflection from lossy chaotic cavities.}

\author{Yan V. Fyodorov}
\affiliation{Department of Mathematics, King's College London, London WC26 2LS, United Kingdom}
\affiliation{L.D.Landau Institute for Theoretical Physics, Semenova 1a, 142432 Chernogolovka, Russia}

\author{Mohammed Osman}

\affiliation{Department of Mathematics, King's College London, London WC26 2LS, United Kingdom}

\begin{abstract}
  Motivated by the phenomenon of Coherent Perfect Absorption, we study the shape of the deepest minima in the frequency-dependent single-channel reflection of waves from a cavity with spatially uniform losses. We show that it is largely determined by non-orthogonality factors $O_{nn}$ of the eigenmodes associated with the non-selfadjoint effective Hamiltonian. For cavities supporting chaotic ray dynamics we then use random matrix theory to derive, fully non-perturbatively, the explicit probability density ${\cal P}(O_{nn})$ of the non-orthogonality factors for systems with both broken and preserved time reversal symmetry. The results imply that $O_{nn}$ are heavy-tail distributed, with the universal tail ${\cal P}(O_{nn}\gg 1)\sim O_{nn}^{-3}$.
\end{abstract}

\maketitle

  A surge of interest related to the proposal of constructing the so-called Coherent Perfect Absorbers (CPA)\cite{Baranov_etal_CPA_2017} has been evident in recent years.  The simplest CPA's can be looked at as a scattering system ( e.g. a cavity)  with a small amount of loss which completely absorbs a monochromatic wave incident at a particular frequency \cite{chong_coherent_2010}. Nonlinear effects may further help to make CPA's a broad-band phenomenon\cite{Suwun_etal_2021} increasing prospects of its practical implementation, including optical filters and switches or logic gates for use in optical computers.
  Recently, a CPA in a rectangular cavity with randomly positioned scatterers and absorption due to a single antenna has been realised experimentally \cite{pichler_random_2019}, paving a way to the construction of CPAs based on disordered cavities. In another recent experiment, a CPA has been realised with a two-port microwave graph system, both with and without time-reversal symmetry \cite{chen_perfect_2020}, and yet in another way with chaotic cavity with programmable meta-atom inclusions \cite{Delhoughne_etal_CPA}. In those settings the relevant framework is that of chaotic wave scattering, and the CPA state corresponds to an eigenstate of the scattering matrix $S(\omega)$ with zero eigenvalue at a real frequency. This fact naturally motivates rising interest in a more general question of characterizing S-matrix complex zeros, as well as their manifestation in physical observables, which has not been systematically studied for wave chaotic systems with losses until very recently \cite{li_random_2017,fyodorov_distribution_2017,Baranov_etal_2017,fyodorov_reflection_2019,OsmanFyodorov2020,Kang_Genak2021}.

  Although CPA's are clearly a new exciting development, a somewhat related phenomena in lossy chaotic scattering have been discussed actually long ago. To this end one may invoke an early influential experiment by Doron, Smilansky and
Frenkel who studied the frequency-dependent reflection coefficient $R(\omega)=|S(\omega)|^2$ of an electromagnetic wave sent via a single-mode waveguide to a cavity shaped in the form of a chaotic billiard. This line of experimental activity maintained its strength over several decades \cite{Kuhletal2003,Hemmadi2005,Hemmadi2006b,microwgraphs2,microwgraphs4}  providing important insights and stimulating theoretical research.

Returning to the setting of \cite{Doron1990} one should recall that the scattering matrix  of a single-channel system without gain or loss must be unimodular due to flux conservation: $S(\omega)=e^{i\delta(\omega)}$, where real $\delta(\omega)$ is known as the scattering phaseshift. The frequency derivative of the scattering phaseshift is an important scattering characteristics called Wigner time delay \cite{wigner55} and is well-described by the sum of Lorentzians of widths $\Gamma_n$ centered at positions $E_n$,
\begin{equation}\label{wigner}
\tau_W(\omega):=\frac{d}{d\omega}\delta(\omega)=2\sum_{n=1}^N\frac{\Gamma_n}{(\omega-E_n)^2+\Gamma_n^2}
\end{equation}
One of the main experimental observations made in \cite{Doron1990} was that the reflection coefficient $R(\omega)$ showed considerable variations with frequency $\omega$, with many pronounced dips to low values
$R(\omega)\lesssim 0.1$ at some frequencies, reminiscent of an ''imperfect'' version of modern CPA. Such a behaviour is clearly incompatible with flux conservation and has been reasonably attributed to presence of uniform losses in resonator walls. Such losses had been then taken into account phenomenologically by adding a small imaginary increment to the real frequency: $\omega\to \omega+i\epsilon$ with $\epsilon>0$.
Assuming that absorption is weak, i.e. $\epsilon\ll \Delta$, where $\Delta$ stands for the mean spacing between eigenfrequencies in the closed cavity in a given frequency range, one may expand in $\epsilon$ yielding a relation between $R(\omega)$ and the Wigner time-delay:
$|S(\omega+i\epsilon)|^2 \approx e^{-2\epsilon \,\tau_W(\omega)}$. The paper thus was among the first
promoting interest in statistics of Wigner time delays in wave-chaotic scattering, which after three decades  still remains an active research topic, see e.g. \cite{Lehmann95,FSS96,FSS97,Brower99,SFS01,OssFyo05,Kott05,MezSimm13,TexMaj013} as well as \cite{Novaes15,Cunden15,Texier16,Grabsch18,Grabsch20,Chen_Anlage_Fyodorov2021} and references therein.
 Measuring the phaseshifts $\delta(\omega)$ independently allowed  to test  experimentally the relation between $R(\omega)$ and the Wigner time-delay, and overall good agreement has been reported in \cite{Doron1990}, with discrepancies close to
the deepest minima attributed to inacuracies in numerical differentiation.

Our goal in this paper is to have a closer look at $R(\omega)$ in the above setting and to demonstrate that its behaviour at
the deepest CPA-like dips bears important physical information which seems to have been not discussed before when addressing
the shape of CPA minima, as e.g. in \cite{Delhoughne_etal_CPA}.
The proper framework for such an analysis is the so called effective Hamiltonian formalism
for wave-chaotic scattering \cite{SokZel89,Fyodorov97,Irotter09,FSav11,kuhl13,Schomerus2015,Fyodorov2016}.
In such a formalism the parameters $E_n$ and $\Gamma_n$ featuring in (\ref{wigner}) describe the positions $z_n=E_n-i\Gamma_n, \, n=1,\ldots, N$ of
$S-$ matrix poles in the lower half of the complex frequency plane, which in turn are considered to be complex eigenvalues
of an $N\times N$ non-selfadjoint matrix  ${\cal H}_{eff}=\mathbf{H}_N-i{\bf w}\otimes {\bf w}^{\dagger}$ known as the effective Hamiltonian. In this setting the eigenfrequencies $\omega_n$ of the closed resonator cavity are associated with the real eigenvalues of the self-adjoint part $\mathbf{H}_N=\mathbf{H}_N^{\dagger}$, with corresponding eigenmodes forming an orthonormal basis. Those eigenfrequencies are converted to the complex $S-$ matrix poles due to coupling of the cavity to
continuum via a single open channel supporting waves coming from (and escaping to) infinity. The coupling between the cavity and the channel is then characterised by the vector ${\bf w}=(w_1,\ldots,w_N)$ of coupling amplitudes.
 The single-channel scattering matrix $S(\omega+i \epsilon)$ of a system with spatially-uniform losses $\epsilon>0$ in this approach can be represented  as
 \begin{equation} \label{Smat1}
 \vspace{-0.2cm}
 S(\omega+i\epsilon)= \prod_{n=1}^{N} \frac{\omega+i\epsilon - z^*_n}{\omega+i\epsilon-z_n},
  \end{equation}
where $z_n^*$ stands for the complex conjugate of $z_n$.
Such relation immediately implies that in the absence of losses $S(\omega)=e^{i\delta(\omega)}$ as dictated by flux conservation, and Eq.(\ref{wigner}) then easily follows. In presence of a uniform absorption $\epsilon>0$ unimodularity is lost, and
  it is completely clear that the deepest dips in $|S(\omega+i\epsilon)|^2$ happen when the condition  $\omega+i\epsilon = E_n+i\Gamma_n$ is approximately satisfied.
  
  \begin{figure}[h!]
    \centering
        \includegraphics[width=0.3\textwidth]{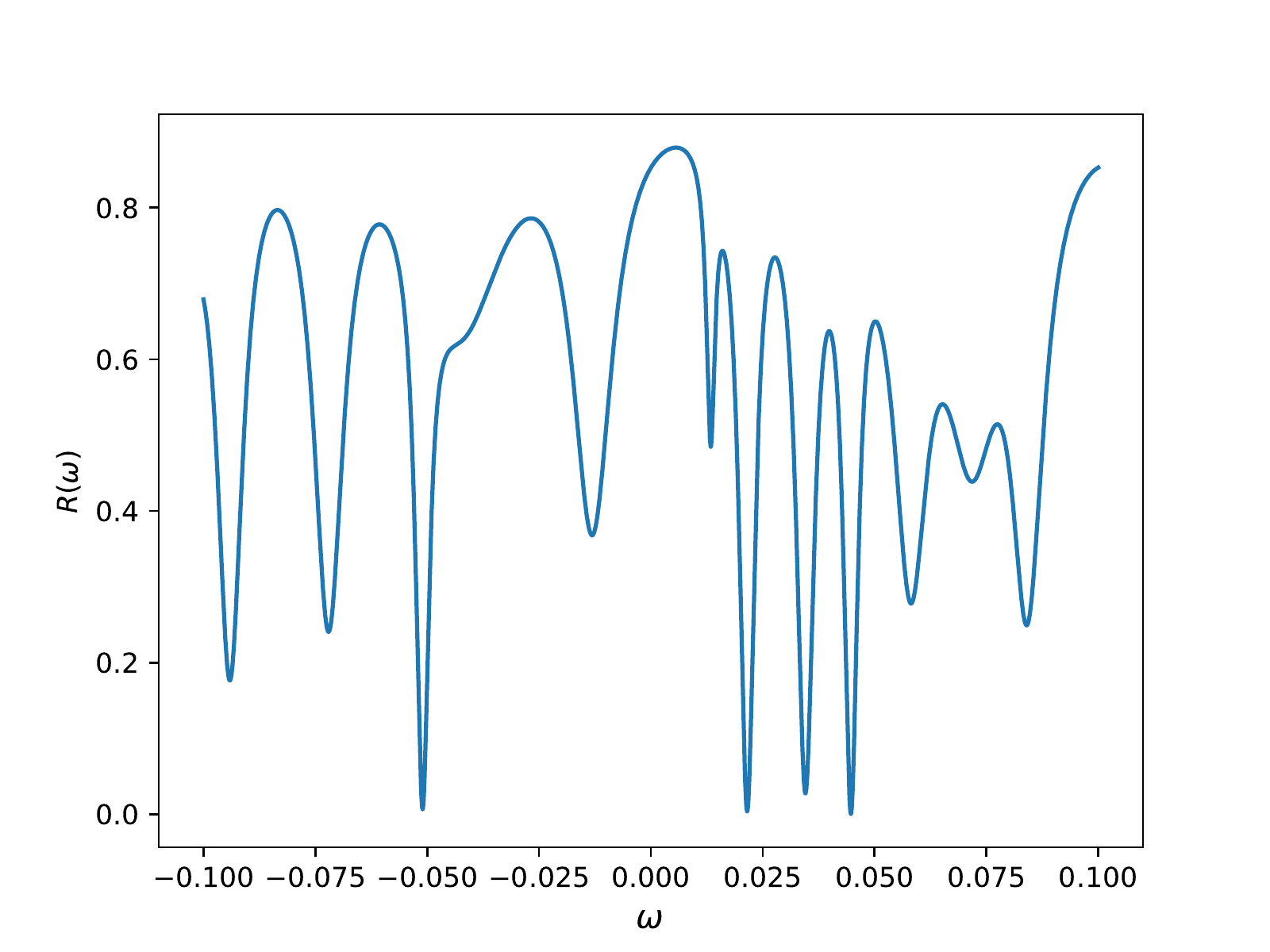}
        \includegraphics[width=0.3\textwidth]{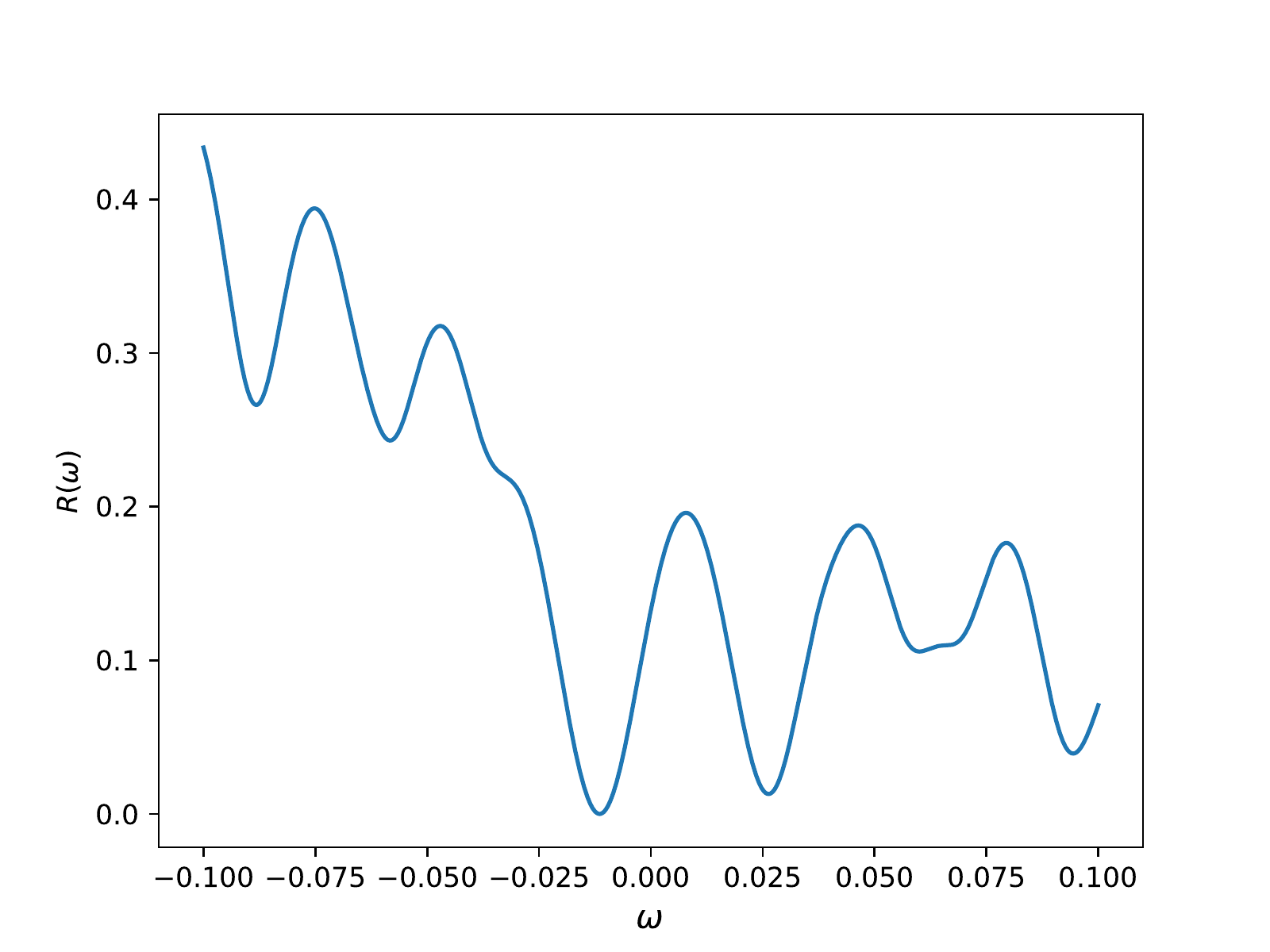}
     \caption{A typical profile of the reflection coefficient at the centre of the spectrum for $N=200,\,\gamma=0.5$ and two different absorptions: $\epsilon=10^{-3}\ll \Delta$ (top) and $\epsilon=0.01\sim \Delta$ (bottom)}
        \label{fig:reflection_profile}
    \end{figure}%
  
  
  It is important to note that for a  single-channel wave-chaotic system the set of parameters $\Gamma_n$ (known as the {\it resonance widths}) is generically random. Note that the value of the uniform loss $\epsilon$ can be reliably extracted from scattering data, as discussed, e.g. in \cite{Baez_etal_2008}. By changing the frequency $\omega$ one always can ensure satisfying $\omega=E_n$ for a given $n$,  but if choosing uniform absorption rate $\epsilon$ is not fully under experimental control the condition $\epsilon=\Gamma_n$ can be satisfied only purely accidentally. If such coincidence however does happen, one comes exactly to the CPA situation of vanishing reflection, see Fig.(\ref{fig:reflection_profile}). In what follows we assume that the reflection dip is located  at the frequency $\omega=E_n$ but may be not perfect, though deep enough, i.e the value at the minimum $0\le R_{min}\ll 1$. This is ensured by imposing the relations $\omega =E_n+ \delta \omega $ and $\Gamma_n=\epsilon+\delta_{n}$ assuming the range of frequencies around the dip $\delta \omega\ll \Delta$ and the resonance mismatch parameter $|\delta_{n}|\ll \epsilon$. Under these conditions using
   Eq.(\ref{Smat1}) we immediately see that the shape of the reflection dip centered at the frequency $\omega=E_n$ should be well described by the following profile:
   \begin{equation}
    \label{profile1}
     R(\omega)=K_n\,\frac{(\omega-E_n)^2+\delta^2_{n}}{(\omega-E_n)^2+4\epsilon^2} , \quad K_n=\prod_{k\ne n}^{N} \frac{|z_k - z_n|^2}{|z^*_k - z_n|^2},
\end{equation}
Such a profile is then uniquely characterized by its depth $R_{min}$ and the curvature at the minimum $C:=\frac{d^2}{d\omega^2} R(\omega)|_{\omega=E_n}$ which are given by
\begin{equation}
\label{profilepar}
     R_{min}=K_n\,\frac{\delta^2_{n}}{4\epsilon^2} , \quad C= \frac{K_n}{2\epsilon^2}.
\end{equation}
The fluctuations of the dip shapes over different reflection minima are thus controlled by the statistics of resonance widths $\Gamma_n$ and that of the factors $K_n$.  Whereas the statistical information about $\Gamma_n$ in a few-channel system is available theoretically \cite{fyod96,fyodorov_systematic_1999,somm99,FyoSav15} and is in a good agreement with experiments \cite{kuhl_resonance_2008},
the factor $K_n$ looks less familiar, and its meaning is not immediately obvious.


  To this end we observe below that such a  factor can be actually given a direct interpretation, by relating it to {\it nonorthogonal eigenfunctions} of the effective Hamiltonian  ${\cal H}_{eff}$. The latter Hamiltonian is not only non-selfadjoint, but also {\it non-normal}, i.e. does not commute with its adjoint, hence is characterized by the pair of left ${\bf l}_n$ and right ${\bf r}_n$ eigenvectors corresponding to the same complex eigenvalue $z_n$, or equivalently ${\cal H}_{eff}{\bf r}_n=z_n{\bf r}_n$ and ${\cal H}^{\dagger}_{eff}{\bf l}_n=z^*_n{\bf l}_n$. One can always choose the sets of right- and left eigenvectors to be {\it bi}-orthogonal, with scalar products satisfying ${\bf l}^{\dagger}_n{\bf r}_m=\delta_{nm}$. The overlap non-orthogonality matrix is then defined as $O_{mn}= ({\bf r}^{\dagger}_m{\bf r}_n)({\bf l}^{\dagger}_n{\bf l}_m)$ and its diagonal entries $O_{nn}= ({\bf r}^{\dagger}_n{\bf r}_n)({\bf l}^{\dagger}_n{\bf l}_n)$ are frequently referred to as self-overlaps.
    For normal matrices ${\bf l}_n={\bf r}_n$ implying $O_{nn}=1, \forall n$, whereas for
 a generic non-normal matrix $O_{nn}>1$  so that the difference $t_n=O_{nn}-1$ can be used as a measure of non-normality.
 It is therefore intriguing to find that the factor $K_n$ featuring in Eq.(\ref{profile1}) turns out to be reciprocal to this simplest characteristics of non-normality:  $K_n=O^{-1}_{nn}$.

 Statistics of non-orthogonal eigenvectors in non-normal random matrices is a topic of active research interest
 in Mathematical and Theoretical Physics starting from the seminal works \cite{CM1998,MC2000}.  Various aspects of non-orthogonality received a lot of attention recently \cite{WaltersStarr2015,Burda2015,Belinchi2017,BD2018,Fyod2018,GW2018,NW2018,Zeit2018,Dubach2019,Metz2019,GNetal2020,Akemann2020a,Akemann2020b,FyoTarn2021}.
In particular, self-overlaps  are controlling sensitivity of the complex eigenvalues to perturbation of matrix entries, and
 $\kappa_n:=\sqrt{O_{nn}}$ are frequently called in the mathematical literature the eigenvalue condition numbers.

To this end, the phenomenon of wave-chaotic scattering provides quite a unique framework for addressing eigenfunction non-orthogonality and its manifestations not only theoretically \cite{SavSok97,jani99,scho00,MehlSant2001,FyoMehl2002,fyodorov_random_2003,pert,FyoSav2012} but also experimentally \cite{gros14,DavyGenack18,DavyGenack19}. In the chaotic scattering context
 self-overlaps are known as Petermann factors, and being responsible for the excess noise
  in lasing resonators have been addressed in an early paper \cite{jani99}. Whereas the mean value for self-overlaps for wave-chaotic scattering in systems with broken time-reversal invariance
   has been calculated nonperturbatively using random Matrix Theory (RMT) techniques in \cite{jani99,FyoMehl2002,fyodorov_random_2003}, any other statistical characteristics of self-overlaps were only so far addressed in the framework of second-order perturbation theory in the limit of weak channel coupling: $\gamma:=|{\bf w}|^2\ll 1$.

   In this paper we develop a method of non-perturbative evaluation of the probability density for the variable $t_n=O_{nn}-1$ valid for the single-channel chaotic scattering at any coupling $\gamma>0$.  For generic non-normal matrices the relations between overlaps
  $O_{nn}$ and complex eigenvalues $z_n$ are very involved and hard to utilize. However in the special case of ${\cal H}_{eff}$ given by a rank-one perturbation of the self-adjoint matrix $\mathbf{H}_N$ such relation is very explicit \cite{FyoMehl2002}, and shows that
  $O_{nn}=K_n^{-1}$, with $K_n$ exactly as defined in Eq.(\ref{profile1}). For completeness we provide a direct verification of this relation in the Supplementary Materials section \cite{SM}.

  We expose main features of our method on a somewhat simpler example of systems  with broken time-reversal invariance, the case of preserved invariance being discussed at the end, with details given in \cite{SM}. We rely on the standard assumption that the $N\times N$ Hermitian matrix ${\bf H}_N$ featuring in the
      definition of effective Hamiltonian belongs to the Gaussian Unitary ensemble of RMT. In the physically relevant limit $N\gg 1$
the corresponding eigenvalue density is then given by $\rho=\frac{1}{2\pi}\sqrt{4-X^2}$, hence the mean eigenvalue spacing $\Delta=1/(N\rho)\sim N^{-1}$.
  The coupling amplitudes $w_i, i=1,\ldots,N$  are normalized in such a way that $\gamma=\sum_i |w_i|^2$ stays finite as $N\to \infty$.
With this choice as $N\gg 1$ statistics of all scattering characteristics varying at frequency scales of the order of $\Delta$ around a spectral point $|X|<2$ can be shown to be universal and depending only on the ``renormalized'' coupling strength $g=\frac{1}{2\pi\rho} \left(\gamma+\frac{1}{\gamma}\right)\ge 1$ but not on the particular choice of $w_i$, see e.g.\cite{SokZel89,VWZ85}. E.g. we can choose ${\bf w}=\sqrt{\gamma}\tilde{{\bf w}}$, with components $\tilde{w}_i$ independent, mean zero complex Gaussian, with   covariance $\left\langle \tilde{w}^*_{i}\tilde{w}_{j}\right\rangle=N^{-1}\delta_{ij}$, where the angular brackets always  denote averaging over relevant distributions. Equivalently, one may choose $\tilde{{\bf w}}$ to be any fixed of unit length, in particular  $\tilde{{\bf w}}={\bf e}$ with ${\bf e}^T=(1,0,\ldots,0)$. Under these assumptions typically $\Gamma_n\sim \Delta$, and we introduce $y_n=\pi \Gamma_n/\Delta$. This is the main parameter controlling both non-Hermiticity and non-normality of the effective Hamiltonian ${\cal H}_{eff}$. As long as $g\gg 1$ the departure from
Hermiticity/normality is small and to a large extent can be studied perturbatively, whereas in the so-called perfect coupling case $g=1$ those effects are the strongest.

With those preparations we can formulate our main result. Define, using Dirac $\delta(x)$ and
$\delta^{(2)}(z):=\delta(Re\, z)\delta(Im\, z)$ the following object
   \begin{equation}
{\cal P}(t;z)=\left\langle\frac{1}{N}\sum_{i=1}^{N}\delta(O_{nn}-1-t)\delta^{(2)}(z-z_n)\right\rangle \label{eq:jointPdfDef}
\end{equation}
interpreted as the (conditional) probability density of the non-orthogonality  factor $t=O_{nn}-1$ corresponding to eigenvalues in the vicinity of a point $z=X-iY, \, Y>0$ in the complex plane. Then
the limiting  density ${\cal P}^{(2)}_y(t): = \lim_{N\to \infty}\frac{1}{\pi\rho N}{\cal P}(t;z=X-i\frac{y}{\pi\rho N})$ takes the following form for the present symmetry class:
\begin{equation}\label{mainTRSB}
{\cal P}^{(2)}_y(t)=\frac{16}{t^3}\,e^{-2gy}\, \mathbb{L}_2\, e^{-2gy\left(1+\frac{2}{t}\right)}I_0\left(\frac{4y}{t}\sqrt{(g^2-1)(1+t)}\right)
\end{equation}
where $I_{\nu}(x)$ stands for the modified Bessel function and $\mathbb{L}_2$ is a differential operator acting on functions $f(y)$ as
\begin{equation}\label{oper} \mathbb{L}_2\, f(y) =\left\{1+\left(\frac{\sinh{2y}}{2y}\right)^2+\frac{1}{2y}\left(1-\frac{\sinh{4y}}{4y}\right)\frac{d}{dy}\right.
\end{equation}
$\left. +\frac{1}{4}\left(\left(\frac{\sinh{2y}}{2y}\right)^2-1\right)\frac{d^2}{dy^2}\right\}\,y^2f(y).$

The following remarks on the above formula are due. First, the most prominent feature in (\ref{mainTRSB}) is the heavy-tail behaviour  ${\cal P}^{(2)}_y(t)\sim t^{-3}$ for $t\gg 1$ rendering all moments $\left\langle O^l_{nn}\right\rangle, \, l\ge 2$ divergent. This tail behaviour is exactly the same as found earlier in other complex-valued non-normal random matrices \cite{BD2018,Fyod2018,Dubach2019} and seems to be the most universal feature of random diagonal overlaps. Second, the integrals $\int {\cal P}^{(2)}_y(t) t^l \,dt$ can be explicitly performed for $l=0$ and $l=1$, reproducing the known mean density of the resonance widths \cite{fyod96} and the mean diagonal overlap  \cite{FyoMehl2002}.
Finally, with suitable adjustments, the method should be applicable to the rank-one family of {\it subunitary} deformations
of the Haar-distributed unitary random CUE matrices. The corresponding model  introduced in \cite{Fyo2000,FyoSom2000} naturally appears
in the context of time-periodic scattering. The statistics of non-orthogonality factors are then expected to be exactly  the same as given by (\ref{mainTRSB}), with the perfect coupling case $g=1$ corresponding in that context to the so-called {\it truncated} CUE \cite{ZS2000}.

Our starting point is to consider (negative) moments of the diagonal overlap defined for $z=X-iY$ and real $p$ as
${\cal O}_p=\left\langle \frac{1}{N}\sum_{n=1}^N\delta^{(2)}(z-z_n)O_{nn}^{-p}\right\rangle$. Fixing the coupling parameter $\gamma>0$ it is technically convenient to consider first the fixed coupling amplitudes ${\bf w}=\sqrt{\gamma}{\bf e}$.
 We note that as $\sum_n Im z_n=-\gamma$ and $Im z_n\le 0, \, \forall n$ it is enough to consider $0\le Y<\gamma$. Following the procedure explained for $p=-1$ in \cite{FyoMehl2002}, we denote $\gamma_*=\gamma-Y>0$,
 use the known joint probability density of all $z_n, \, n=1,\ldots, N$ \cite{fyodorov_systematic_1999,StSe98,Kozhan07}, and rely on  GUE invariance. After defining $\tilde{z}=\sqrt{\frac{N}{N-1}}z$, considering
$(N-1)\times (N-1)$ matrices $\tilde{{\cal H}}_{eff}:=\mathbf{H}_{N-1}-i\sqrt{\frac{N}{N-1}}\gamma_*{\bf e}_{N-1}\otimes {\bf e}_{N-1}^T $
and disregarding the explicit proportionality constants we arrive at the following representation
\begin{equation}\label{Mom1GUE}
{\cal O}_p(z)\propto \frac{\gamma_*^{N-2}}{\gamma^{N-1}}\,e^{-\frac{N}{2}(X^2-Y^2+\gamma^2)+\frac{N-1}{2}\gamma_*^2} {\cal M}^{(\beta=2)}_p(z)
\end{equation}
where
\begin{equation}\label{Mom2}
{\cal M}^{(\beta=2)}_p(z)=\left\langle\frac{\left[\det(\tilde{z}-\tilde{{\cal H}}_{eff})\det(\tilde{z}^*-\tilde{{\cal H}}^{\dagger}_{eff})\right]^{p+1}}{\left[\det(\tilde{z}-\tilde{{\cal H}}^{\dagger}_{eff})\det(\tilde{z}^*-\tilde{{\cal H}}_{eff})\right]^p}\right\rangle
\end{equation}
 with the average now going over the reduced  GUE matrices $\mathbf{H}_{N-1}$ of the size $(N-1)\times (N-1)$. The major task is to evaluate
${\cal M}^{(\beta=2)}_p(z)$ as $N\to \infty$ for a given $\gamma>0$ and real $p$  assuming the scaling $Im z=Y\sim N^{-1}$. It is easy to check  that in such a limit one may safely replace $\tilde{{\cal H}}_{eff}$ with the original ${\cal H}_{eff}=\mathbf{H}_N-i\gamma\,\tilde{{\bf w}}\otimes \tilde{{\bf w}}^{\dagger}$, without changing the result. When evaluating ${\cal M}^{(\beta=2)}_p(z)$ we also find it more technically convenient  to consider the  random complex gaussian channel vector ${\bf \tilde{w}}$ rather than keeping it of unit length, and can show that both choices lead to the same result.

The crucial trick which allows to evaluate  ${\cal M}^{(\beta=2)}_p(z)$  efficiently is to define a version of the matrix element of the GUE resolvent (Green's function), also known in scattering theory as the $K-$matrix, via
\begin{equation} \label{reso}
G(z):={\bf \tilde{w}}^{\dagger}\frac{1}{z-\mathbf{H}_{N}}{\bf \tilde{w}}=u+iv, \, v>0
\end{equation}
and to use the identity $\det(z-{\cal H}_{eff})=\det(z-\mathbf{H}_{N})(1+i\gamma G(z))$
to rewrite our  main object of interest as
 \begin{equation}\label{Mom3}
{\cal M}^{(\beta=2)}_p(z)=\left\langle |\det(z-\mathbf{H}_{N})|^2\frac{\left|1+i\gamma G(z)\right|^{2(p+1)}}{\left|1+i\gamma G(z^*)\right|^{2p}}\right\rangle
\end{equation}
 \begin{equation}\label{Mom4}
=\int_{-\infty}^{\infty}du\int_0^{\infty}dv \frac{\left[(1-\gamma v)^2+\gamma^2 u^2\right]^{p+1}}{\left[(1+\gamma v)^2+\gamma^2 u^2\right]^p}{\cal P}^{(\beta=2)}_{z}(u,v)
\end{equation}
in terms of the function
\begin{equation}\label{PjpdGUE}
{\cal P}^{(\beta=2)}_{z}(u,v)=\left\langle |\det(z-\mathbf{H}_{N})|^2 \right.
\end{equation}
\[
\left.\times \delta\left(u-\mbox{Re} G(z)\right)\,\delta\left(v-\mbox{Im} G(z)\right) \right\rangle
\]
where now averaging includes one over $\tilde{{\bf w}}$ as well. Such function represents a ``deformed'' version of
the probability density for the Green's function entry, Eq.(\ref{reso}).
Without deformation it enjoyed thorough attention, starting from \cite{MF1994}, and methods to evaluate it within RMT context (and beyond)  has been discussed e.g. in \cite{fyodsav04,savfyodsom05}.
Following one of those methods, we consider the Fourier-Laplace transform
${\cal F}(s,p)=\int du \frac{e^{isu}}{2\pi}\int_0^{\infty} dv e^{-pv} {\cal P}_z(u,v)$, and perform the average over
 $\tilde{{\bf w}}$ first, before that over the GUE matrices $\mathbf{H}_{N}$.
Along those lines we notice that
$$is \mbox{Re}\, G(z) - p\mbox{Im}\, G(z)={\bf
\tilde{ w}}^{\dagger}\frac{is(X-\mathbf{H}_{N})-pY}{(z-\mathbf{H}_{N})(\overline{z}-\mathbf{H}_{N})}{\bf \tilde{w}}$$
so that the average over ${\bf \tilde{w}}$ amounts to performing a simple Gaussian integral. For simplicity of the presentation we set $X=0$ below, the results for any $|X|<2$ recovered by a simple rescaling, and scale $Y=iy/N$, Doing this we arrive after simple algebraic manipulations to the following identity
\begin{equation}\label{Mom5}
 {\cal F}^{(\beta=2)}(s,p)=\left\langle\frac{|\det(\frac{i y}{N}-\mathbf{H}_{N})|^4}
 {\det\left(-i\frac{\tau_{+}}{2N}-\mathbf{H}_{N}\right)\det\left(-i\frac{\tau_{-}}{2N}-\mathbf{H}_{N}\right)}\right\rangle
\end{equation}
where  we defined  $\tau_{+}=s+\sqrt{s^2+4y(y+p)}>0$ and  $\tau_{-}=s-\sqrt{s^2+4y(y+p)}<0$.
 The objects like those featuring in the right-hand side of (\ref{Mom5})
have a long history of studies, starting from \cite{AS1995}, and then in increasing generality in \cite{FyoStra,StraFyo}
culminating in \cite{BorStra}. Using the latter results an explicit expression for ${\cal F}(s,p)$ as $N\to \infty$ can be found as explained in   \cite{SM}. Then on performing the Laplace-Fourier inversion of ${\cal F}(s,p)$  \cite{SM} one arrives at a quite compact result:
\begin{equation}\label{jpd}
{\cal P}^{(\beta=2)}_{z}(u,v)\propto  \mathbb{L}_2 \frac{1}{v^3} e^{-y\frac{u^2+v^2+1}{v}}
\end{equation}
 with the differential operator $ \mathbb{L}_2$ defined in (\ref{oper}).
Now we can substitute  (\ref{jpd}) back to  (\ref{Mom4}) and after introducing a new variable $x=\frac{u^2+v^2+1}{2v}>1$ arrive to
the formula for $M^{(\beta=2)}_p(y)=\lim_{N\to \infty}{\cal M}^{(\beta=2)}_p(z=-\frac{iy}{N})$ \cite{SM}:
\begin{equation}\label{Mom7}
 M^{(\beta=2)}_p(y)\propto \gamma \mathbb{L}_2\int_1^{\infty}da \frac{(a-1)^{p+1}}{(a+1)^p}
\end{equation}
\[
\times e^{-2yga}I_0(2y\sqrt{(a^2-1)(g^2-1)})
\]
The above formula is well-defined for any $p>-2$.
Evaluating straightforwardly the limit $N\to \infty$ in (\ref{Mom1GUE}) while keeping $X=0, Y=y/N$
gives us the moments of the self-overlap as ${\cal O}_p\propto e^{-2gy}M_p(y)$ and the main result  (\ref{mainTRSB}) then follows by a straightforward moment inversion after setting $a=2t^{-1}+1$.

The same method can be adapted {\it mutatis mutandis} for deriving a similar result for systems with preserved time-reversal symmetry, with the matrix $\mathbf{H}_{N}$ belonging to the Gaussian Orthogonal Ensemble of RMT and the channel coupling vector being real-valued\cite{footnote}. In that setting the central result needed for our goal is a correlation function involving {\it half-integer} powers of characteristic polynomials evaluated in \cite{FyoNock2015}. Necessary steps of the derivation are presented in \cite{SM}, with the final result being now
 \begin{equation}\label{mainTRS}
{\cal P}^{(1)}_y(t)=\frac{1}{2}\frac{\,e^{-gy}}{\sqrt{t^5(1+t)}}\,\, \mathbb{L}_1\, e^{-gy\left(1+\frac{2}{t}\right)}I_0\left(\frac{2y}{t}\sqrt{(g^2-1)(1+t)}\right)
\end{equation}
where $\mathbb{L}_1$ is the following differential operator
\begin{equation} \mathbb{L}_1 =2\sinh{2y}-\left(\cosh{2y}-\frac{\sinh{2y}}{2y}\right)\left(\frac{3}{y}+2\frac{d}{dy}\right)
\end{equation}
We see that  the heavy-tail behaviour  ${\cal P}^{(1)}_y(t)\sim t^{-3}$ for $t\gg 1$ is also manifest in such a case. One can further integrate over $t$ to obtain the mean density of the scaled resonance widths $y=\pi\Gamma_n/\Delta$ in the form:
 \begin{equation}\label{densityTRS}
\rho^{(1)}(y)=\frac{1}{4\sqrt{2}}e^{-gy}\,\, \mathbb{L}_1 \Phi(y)
\end{equation}
where
 \begin{equation}\label{densityTRS}
 \Phi(y) = \int_1^{\infty}da\,e^{-gay}\frac{(a-1)}{\sqrt{a+1}}I_0\left(y\sqrt{(g^2-1)(a^2-1)}\right)
 \end{equation}
This expression looks considerably simpler than the 3-fold integral representation derived earlier for the same density in \cite{{somm99}}, though one can check numerically that they are fully equivalent \cite{SM}.

In conclusion, we related the shape of deep, CPA-like dips in a single-channel reflection from lossy cavities
with the diagonal non-orthogonality factor of eigenfunctions of the underlying effective Hamiltionian. Using RMT framework we derived, fully non-perturbatively, the explicit distribution of  these factors for wave-chaotic scattering in systems with both broken and preserved time reversal symmetry. The results imply that $O_{nn}$ are heavy-tail distributed, sharing this feature with other instances of non-orthogonality factors of non-Hermitian ensembles \cite{BD2018,Fyod2018,Dubach2019,FyoTarn2021} and supporting the claim of the universality of such a behaviour. Experimentally, statistics of $O_{nn}$ should be accessible within the framework of ''harmonic inversion'' method \cite{kuhl_resonance_2008}, or via accurate study of the shape of reflection dips.

Y.V.F. acknowledges financial support from EPSRC Grant EP/V002473/1 ''Random Hessians and Jacobians: theory and applications''.

\newpage
 \begin{widetext}

\section{SUPPLEMENTARY MATERIALS}

\subsection{Non-orthogonality factors in the single-channel effective Hamiltonian}

Recall the definitions of the effective Hamiltonian ${\cal H}_{eff}=\mathbf{H}_N-i{\bf w}\otimes {\bf w}^{\dagger}$
 and its left-right eigenvectors and eigenvalues:
\begin{equation} \label{defLR}
{\cal H}_{eff}{\bf r}_n=z_n{\bf r}_n, \,\, {\bf l}^{\dagger}_n{\cal H}_{eff}=z_n{\bf l}^{\dagger}_n  \quad \mbox{ and }
{\bf r}^{\dagger}_n{\cal H}^{\dagger}_{eff}=z^*_n{\bf r}^{\dagger}_n \,\,\quad {\cal H}^{\dagger}_{eff}{\bf l}_n=z^*_n{\bf l}_n\,.
\end{equation}
We choose the sets of right- and left eigenvectors to be {\it bi}-orthogonal, with scalar products satisfying
${\bf l}^{\dagger}_n{\bf r}_m=\delta_{nm}$.

In these terms the single-channel scattering matrix at the complex plane $z=\omega+i\epsilon$ can be rewritten, after a simple algebra,  in several equivalent forms
\begin{equation} \label{Smat2}
    S(z)=\prod_{n=1}^{N} \frac{z - z^*_n}{z-z_n}=  \frac{\det{\left(z -{\cal H}_{eff}^{\dagger}\right)} }{\det{\left(z -{\cal H}_{eff}\right)}}
 \end{equation}
 \begin{equation} \label{Smat3}
   =1-2i{\bf w} ^{\dagger}\frac{1}{z-{\cal H}_{eff}}{\bf w}=1-2i\sum_{n=1}^N \frac{\left({\bf w} ^{\dagger}{\bf r}_n\right)\left({\bf l}^{\dagger}_n{\bf w}\right)}{z-z_n}
 \end{equation}
 Comparing the residues at the poles $z_n$ in  (\ref{Smat2}) and (\ref{Smat3}) one finds the relation:
 \begin{equation} \label{Smat4}
   -2i\left({\bf w} ^{\dagger}{\bf r}_n\right)\left({\bf l}^{\dagger}_n{\bf w}\right)=(z_n-z_n^*)\prod_{k\ne n}\frac{z_k-z_n^*}{z_k-z_n}
 \end{equation}
 Conjugation gives also:
  \begin{equation} \label{Smat5}
   2i\left({\bf w} ^{\dagger}{\bf l}_n\right)\left({\bf r}^{\dagger}_n{\bf w}\right)=(z^*_n-z_n)\prod_{k\ne n}\frac{z^*_k-z_n}{z^*_k-z^*_n}
 \end{equation}
 Multiplying (\ref{Smat4}) and (\ref{Smat5}) and rearranging gives:
 \begin{equation} \label{Smat6}
   \left[2i\left({\bf l}^{\dagger}_n{\bf w}\right)\left({\bf w} ^{\dagger}{\bf l}_n\right)\right]\left[-2i\left({\bf r}^{\dagger}_n{\bf w}\right)\left({\bf w} ^{\dagger}{\bf r}_n\right)\right]=-(z_n-z_n^*)^2\prod_{k\ne n}\frac{|z^*_k-z_n|^2}{|z_k-z_n|^2}
 \end{equation}

On the other hand one may notice from the definition that $2i{\bf w} \otimes {\bf w} ^{\dagger}= {\cal H}_{eff}^{\dagger}-{\cal H}_{eff}$, hence
\begin{equation} \label{Smat7}
   2i\left({\bf l}^{\dagger}_n{\bf w}\right)\left({\bf w} ^{\dagger}{\bf l}_n\right)={\bf l}^{\dagger}_n \, \left[  2i{\bf w}\otimes {\bf w} ^{\dagger}\right]\,{\bf l}_n={\bf l}^{\dagger}_n \, \left[ {\cal H}_{eff}^{\dagger}-{\cal H}_{eff} \right]\,{\bf l}_n=(z_n^*-z_n)
   \left({\bf l}^{\dagger}_n{\bf l}_n\right)
 \end{equation}
 and similarly:
 \begin{equation} \label{Smat8}
   -2i\left({\bf r}^{\dagger}_n{\bf w}\right)\left({\bf w} ^{\dagger}{\bf r}_n\right)={\bf r}^{\dagger}_n \, \left[  -2i{\bf w}\otimes {\bf w} ^{\dagger}\right]\,{\bf r}_n={\bf r}^{\dagger}_n \, \left[ {\cal H}_{eff}-{\cal H}_{eff}^{\dagger} \right]\,{\bf r}_n=(z_n-z_n^*)
   \left({\bf r}^{\dagger}_n{\bf r}_n\right)
\end{equation}
Substituting back to (\ref{Smat5}) shows that:
\begin{equation} \label{Smat9}
\left({\bf l}^{\dagger}_n{\bf l}_n\right) \left({\bf r}^{\dagger}_n{\bf r}_n\right) =  \prod_{k\ne n}\frac{|z^*_k-z_n|^2}{|z_k-z_n|^2}
 \end{equation}
Finally recalling the definition of overlap non-orthogonality matrix as $O_{mn}= ({\bf r}^{\dagger}_m{\bf r}_n)({\bf l}^{\dagger}_n{\bf l}_m)$ we see that the right-hand side of (\ref{Smat9}) provides the expression for diagonal entries $O_{nn}$, and comparing with the definition Eq.(3) of the factor $K_n$ we see that the two are reciprocal.

Finally note that as residues of the scattering matrix at resonance poles can be extracted from the same harmonic inversion procedure used to get experimental access to widths $\Gamma_n$, the relation (\ref{Smat4}) could provide a basis for experimentally studying 
the statistics of the non-orthogonality factors. 

\subsection{Correlation function of characteristic polynomials for GUE matrices, eq.(14)}

For convenience of the reader we collect the formulas from Theorem 1.3.2  in [1] below most useful for our purposes.
Denoting the limiting mean eigenvalue density of $N\times N$ GUE matrices $\mathbf{H}_{N}$ to be $\rho=\frac{1}{2\pi}\sqrt{4-X^2}$ we define $D(\alpha)=\det{\left(X+\frac{\alpha}{N\rho(X)}-\mathbf{H}_{N}\right)}$ and keep $\alpha$ fixed as $N\to \infty$. Then in such a limit
\begin{equation}\label{corrpol1}
\left\langle \frac{D\left(\alpha^{-}_1\right)D\left(\alpha^{-}_2\right)D\left(\beta^{-}_1\right)D\left(\beta^{-}_2\right)}{D(\alpha_{+})D(\beta_{+})} \right\rangle\propto e^{N\frac{X^2}{2}+\frac{X}{\rho(x)}} \frac{\left(\alpha_1^{-}-\alpha_{+}\right)\left(\alpha_2^{-}-\alpha_{+}\right)\left(\beta_1^{-}-\beta_{+}\right)\left(\beta_2^{-}-\beta_{+}\right)}
{\left(\alpha_1^{-}-\alpha_2^{-}\right)\left(\beta_1^{-}-\beta_2^{-}\right)}\,\det{{\cal D}}
\end{equation}
where ${\cal D}$ is the following $3\times 3$ matrix:
\begin{equation}\label{corrpol2}
{\cal D}=\left(\begin{array}{ccc}S(\alpha_1^{-},\beta_1^{-}) & S(\alpha_1^{-},\beta_2^{-}) &S(\alpha_1^{-},\alpha_{+})\\
S(\alpha_2^{-},\beta_1^{-}) & S(\alpha_2^{-},\beta_2^{-}) &S(\alpha_2^{-},\alpha_{+})\\S(\beta_{+},\beta_1^{-}) & S(\beta_{+},\beta_2^{-}) &S(\beta{+},\alpha_{+})
\end{array}\right)
\end{equation}
where in our particular case:
\[\alpha_{1,2}^{-}=i\pi \rho(X) y_{1,2}, \quad \beta_{1,2}^{-}=-i\pi \rho(X) y_{1,2},\quad \alpha_{+}=-i\frac{\pi \rho(X)}{2}\tau_{-}, \quad \beta_{+}=-i\frac{\pi \rho(X)}{2}\tau_{+}  \]
 As $\mbox{Im}\alpha_{+}>0, \, \mbox{Im}\beta_{+}<0$ we have:
 \begin{equation}\label{corrpol3}
S(\alpha_{1}^{-},\beta_{1}^{-})=\frac{1}{\pi}\frac{\sin{\left(\alpha_{1}^{-}-\beta_{1}^{-}\right)}}{\alpha_{1}^{-}-\beta_{1}^{-}}
=\frac{1}{\pi}\frac{\sin{2\pi \rho(X)y_{1}}}{2\pi \rho(X)y_{1}}
\end{equation}
and similarly
\begin{equation}\label{corrpol4}
S(\alpha_{2}^{-},\beta_{2}^{-})=\frac{1}{\pi}\frac{\sin{2\pi \rho(X)y_{2}}}{2\pi \rho(X)y_{2}}, \quad S(\alpha_{1}^{-},\beta_{2}^{-})=S(\alpha_{2}^{-},\beta_{1}^{-})=
\frac{1}{\pi}\frac{\sin{2\pi \rho(X)(y_1+y_2)}}{2\pi \rho(X)(y_1+y_2)}.
\end{equation}
Further
\begin{equation}\label{corrpol5}
S(\beta_{+},\beta_{1,2}^{-})=\frac{e^{\pi \rho(X)\left(y_{1,2}-\frac{\tau_+}{2}\right)}}{i\pi \rho(X)\left(y_{1,2}-\frac{\tau_+}{2}\right)}, \quad S(\alpha_{1,2}^{-},\alpha_{+})=\frac{e^{\pi \rho(X)\left(y_{1,2}+\frac{\tau_{-}}{2}\right)}}{i\pi \rho(X)\left(y_{1,2}+\frac{\tau_{-}}{2}\right)}
\end{equation}
and finally
\begin{equation}\label{corrpol6}
S(\beta_{+},\alpha_{+})=-2\pi i \frac{e^{\alpha_{+}-\beta_{+}}}{\alpha_{+}-\beta_{+}}=2\pi
\frac{e^{\frac{\pi \rho(X)}{2}\left(\tau_{-}-\tau_+\right)}}{\frac{\pi \rho(X)}{2}\left(\tau_{-}-\tau_+\right)}, \quad
\end{equation}

Eventually we need to consider the limit $y_1\to y_2=y$ in (\ref{corrpol1}) which after straightforward manipulations produces (after setting $X=0$, hence $\pi\rho(x)=1$)
the following expression for the Laplace-Fourier transform defined in Eq.(14) of the main paper:
\begin{equation}
{\cal F}^{(\beta=2)}(s,p)=A(s,p)+B(s,p)+C(s,p)
\end{equation}
where
\begin{equation}\label{A}
A(s,p)=2\frac{e^{\frac{1}{2}(\tau_{-}-\tau_{+})}}{\frac{1}{2}(\tau_{-}-\tau_{+})}
\left[\left(y+\frac{\tau_{-}}{2}\right)\left(y-\frac{\tau_{+}}{2}\right)\right]^2
\, \frac{1}{4y^2}\left(\frac{\sinh^2{2y}}{4y^2}-1\right),
\end{equation}

\begin{equation}\label{B}
B(s,p)=\frac{1}{2}e^{\frac{1}{2}(\tau_{-}-\tau_{+})}
\left(y+\frac{\tau_{-}}{2}\right)\left(y-\frac{\tau_{+}}{2}\right)
\, \frac{1}{4y^3}\left(e^{4y}-1-4y-8y^2\right),
\end{equation}
and
\begin{equation}\label{C}
C(s,p)=\frac{1}{2}e^{\frac{1}{2}(\tau_{-}-\tau_{+})}
\left(2y+\frac{\tau_{-}-\tau_{+}}{2}\right) \frac{1}{4y^2}\left(1+4y-e^{4y}\right)
+\frac{1}{2}e^{\frac{1}{2}(\tau_{-}-\tau_{+})}
 \frac{1}{2y}\left(e^{4y}-1\right).
\end{equation}

\subsubsection{Inverting the Laplace-Fourier transform}

Substituting the definitions
\begin{equation}
\tau_{+}=s+\sqrt{s^2+4y(y+p)} \quad \mbox{and} \quad  \tau_{-}=s-\sqrt{s^2+4y(y+p)}
\end{equation}
to the expression  for the Fourier-Laplace transform $F^{(\beta=2)}(s,p)$ presented in Eqs.(\ref{A})-(\ref{C}) above, we notice that it can be represented as
\begin{equation}\label{A1}
A(s,p)=2\frac{e^{-\sqrt{s^2+4y(y+p)}}}{\sqrt{s^2+4y(y+p)}}
\left(2y+p-\sqrt{s^2+4y(y+p)}\right)^2
 \frac{1}{4}\left(\frac{\sinh^2{2y}}{4y^2}-1\right),
\end{equation}
\begin{equation}\label{B1}
B(s,p)=\frac{1}{2}e^{-\sqrt{s^2+4y(y+p)}}
\left(2y+p-\sqrt{s^2+4y(y+p)}\right)
 \frac{1}{4y^2}\left(e^{4y}-1-4y-8y^2\right),
 \end{equation}
 and
\begin{equation}\label{C1}
C(s,p)=\frac{1}{2}e^{-\sqrt{s^2+4y(y+p)}}\left\{
\left(2y-\sqrt{s^2+4y(y+p)}\right) \frac{1}{4y^2}\left(1+4y-e^{4y}\right)+\frac{1}{2y}\left(e^{4y}-1\right)\right\}.
\end{equation}
To represent above as an explicit Laplace-Fourier transform we define for $p>0$, any real $s$ and  $\alpha$ the functions:
\begin{equation}\label{FL1}
\phi_{\alpha}(s,p;y)=\int_0^{\infty}dv e^{-pv}\int_{-\infty}^{\infty}\frac{du}{\sqrt{2\pi}}e^{isu}\frac{1}{v^{\alpha+2}} e^{-y\frac{u^2+v^2+1}{v}}=\frac{1}{2^{\alpha}\,y^{\alpha+1}}\left(s^2+4y(y+p)\right)^{\frac{1}{2}
\left(\alpha+\frac{1}{2}\right)}\,K_{\alpha+\frac{1}{2}}
\left(\sqrt{s^2+4y(y+p)}\right)
\end{equation}
where we performed first the Gaussian integral over $u$ and then used the formula
\[
\int_0^\infty\,dv \, v^{\nu-1}\,e^{-\frac{\beta}{v}-\alpha v}=2\left(\frac{\beta}{\alpha}\right)^{\nu/2}K_{\nu}(2\sqrt{\beta\alpha}), \quad \beta>0, \alpha>0
\]
remembering $K_{\nu}(x)=K_{-\nu}(x)$.

In particular:
\begin{equation}\label{FL2}
 \phi_{1}(s,p;y)=\sqrt{\frac{\pi}{2}}\frac{e^{-\sqrt{s^2+4y(y+p)}}}{2y^2}\left(1+\sqrt{s^2+4y(y+p)}\right),
\end{equation}
and using this we can easily check that
\begin{equation}\label{FL3}
A(s,p)=
 \frac{1}{2}\left(\frac{\sinh^2{2y}}{4y^2}-1\right)\left(\frac{1}{2}\frac{d^2}{dy^2}+2\frac{d}{dy}+2\right)\sqrt{\frac{2}{\pi}}y^2\phi_1(s,p;y),
\end{equation}
Further, after rewriting
\begin{equation}\label{BC}
B(s,p)+C(s,p)=\frac{1}{2}e^{-\sqrt{s^2+4y(y+p)}}\,
(2y+p) \frac{1}{4y^2}\left(e^{4y}-1-4y-8y^2\right)+e^{-\sqrt{s^2+4y(y+p)}}\left(1+\sqrt{s^2+4y(y+p)}\right),
 \end{equation}
it turns out to be possible to represent it in the form
\begin{equation}\label{FL4}
B(s,p)+C(s,p)=
\left(2-\frac{1}{8y^2}\left(e^{4y}-1-4y-8y^2\right)\frac{d}{dy}\right)\sqrt{\frac{2}{\pi}}y^2\phi_1(s,p;y),
\end{equation}
Finally, one may check that adding (\ref{FL3}) and (\ref{FL4}) produces, after simple algebra, eq.(15) of the main text.

\subsection{Derivation of eq.(16) of the main text.}
According to Eqs.(12) and (15) of the main text we need to deal with the following integral:
\begin{equation}\label{Mom4SM}
F^{(2)}_p(y)=\int_{-\infty}^{\infty}du\int_0^{\infty}dv \frac{\left[(1-\gamma v)^2+\gamma^2 u^2\right]^{p+1}}{\left[(1+\gamma v)^2+\gamma^2 u^2\right]^p} \frac{1}{v^3} e^{-y\frac{u^2+v^2+1}{v}}
\end{equation}
which after introducing the combination $x=\frac{1}{2}\left(v+\frac{1}{v}\right)+\frac{u^2}{2v}>1$ as the integration variable instead of $u^2$ allows to rewrite the above as
\begin{equation}\label{Mom4SMa}
F_p(y)\propto \int_{1}^{\infty}dxe^{-2x}\int_{v_1}^{v_2}\frac{dv}{v} \frac{1}{\sqrt{(v-v_1)(v_2-v)}}
 \frac{\left[\frac{1-\gamma^2}{2v}-\gamma +\gamma^2x\right]^{p+1}}{\left[\frac{1-\gamma^2}{2v}+\gamma +\gamma^2x\right]^p}
\end{equation}
where $v_{1,2}=x\mp \sqrt{x^2-1}$, with $v_1v_2=1$. Changing $v\to 1/v$ and then substituting $v=\frac{1}{2}\left(v_1+v_2\right)
+ \frac{1}{2}\left(v_2-v_1\right)\cos{\theta}$, where $\theta \in [0,\pi]$ brings the right-hand side to
\begin{equation}\label{Mom4SMb}
F_p(y)\propto \int_{1}^{\infty}dxe^{-2x}\int_0^{\pi}
 \frac{\left[x(1+\gamma^2)-2\gamma +(1-\gamma^2)\sqrt{x^2-1}\cos{\theta}\right]^{p+1}}{\left[x(1+\gamma^2)+2\gamma +(1-\gamma^2)\sqrt{x^2-1}\cos{\theta}\right]^p}\,d\theta
\end{equation}
 After introducing $g=\frac{1}{2}\left(\gamma+\frac{1}{\gamma}\right)$
and using that $\frac{1}{2}\left(\frac{1}{\gamma}-\gamma\right)=\pm \sqrt{g^2-1}$ with sign dependent on whether $\gamma<1$ or $\gamma>1$
(this sign is immaterial for the value of the integral and can be omitted) the above is brought to the following form:
\begin{equation}\label{Mom4SMc}
F_p(y)\propto \gamma \int_{1}^{\infty}dxe^{-2x}\int_0^{\pi}d\theta \frac{[gx-1+\cos{\theta}\sqrt{(g^2-1)(x^2-1)}]^{p+1}}{[gx+1+\cos{\theta}\sqrt{(g^2-1)(x^2-1)}]^{p}}
 \end{equation}
Further introducing $a=gx+\cos{\theta}\sqrt{(g^2-1)(x^2-1)}>1$ as a new variable one can transform (\ref{Mom4SMc})  to
\begin{equation}\label{Mom4SMd}
 F_p(y)\propto \gamma \int_1^{\infty}da \frac{(a-1)^{p+1}}{(a+1)^p} \int_{x_1(a)}^{x_2(a)}\,e^{-2yx}\frac{dx}{\sqrt{(x-x_1(a))(x_2(a)-x)}}
\end{equation}
$x_{1,2}(a)=ga\mp \sqrt{(a^2-1)(g^2-1)}$. Substituting $x=\frac{1}{2}\left(x_1(a)+x_2(a)\right)
+ \frac{1}{2}\left(x_2(a)-x_1(a)\right)\cos{\phi}$, where $\phi \in [0,\pi]$ brings the right-hand side to
\begin{equation}\label{Mom4SMd}
 F_p(y)\propto \gamma \int_1^{\infty}da \frac{(a-1)^{p+1}}{(a+1)^p} e^{-2gay} \int_{0}^{\pi}\,e^{-2y\sqrt{(a^2-1)(g^2-1)}\cos{\phi}}\,d\phi
\end{equation}
which is equivalent to (16) of the main text.

 Numerical simulations provide a reasonably good support for the final result,  Eq.(6) of the main text, already for relatively small matrices.
  \begin{figure}[h!]
        \includegraphics[width=0.5\textwidth,keepaspectratio=true]{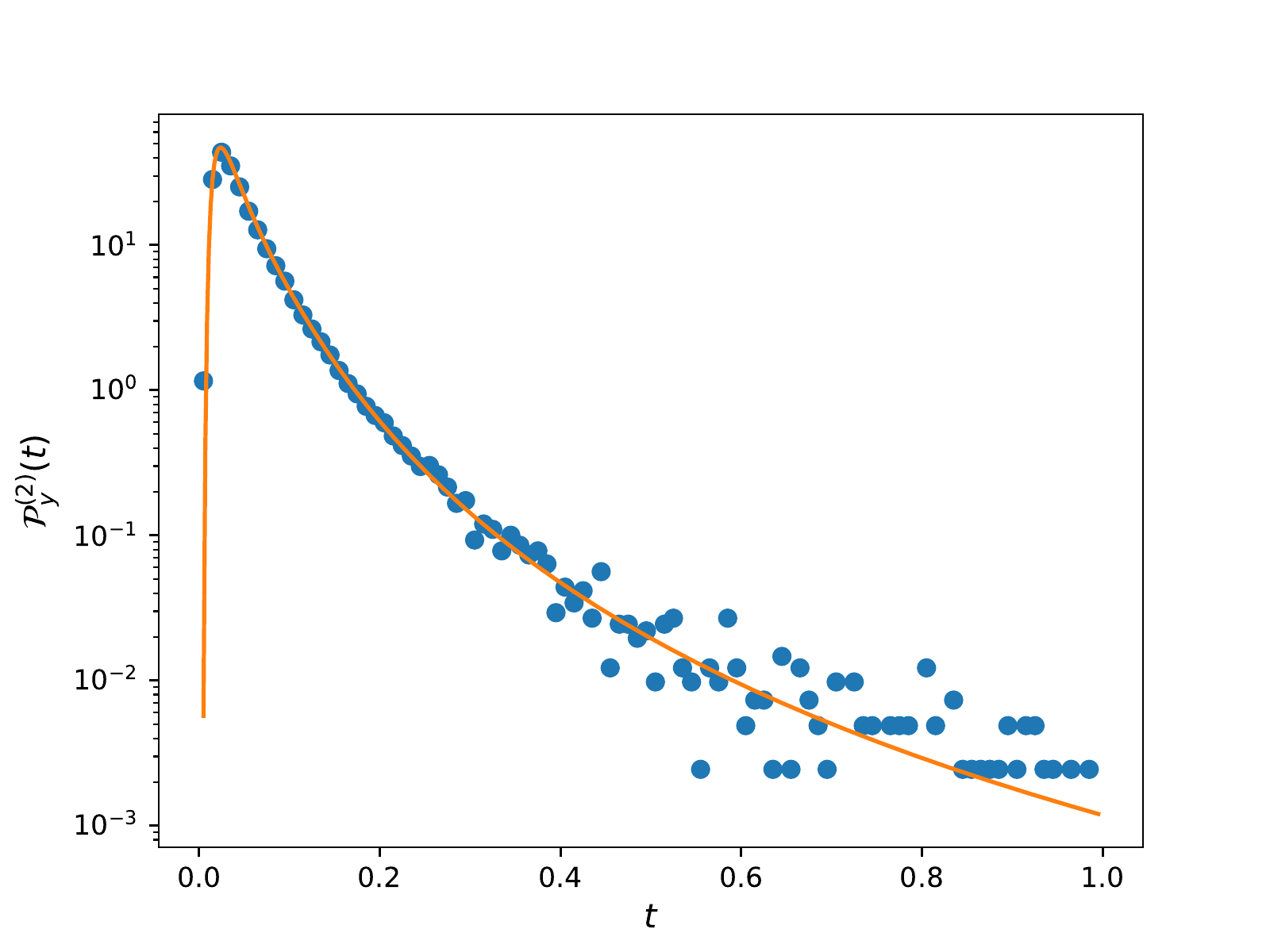}
        \label{fig:beta=2y=0.05}
     \caption{ Eq.6 compared with the empirical density of overlaps $O_{nn}$ for $n=10^{6}$ realizations of $32\times32$ GUE matrices, with the choice $y=0.05$ and $\gamma=0.5$.}
\end{figure}

\subsection{Non-orthogonality factors in systems with preserved time-reversal invariance}
Our starting point is the well-known joint probability density (JPD) of eigenvalues for $N\times N$ non-selfadjoint matrix  ${\cal H}_{eff}=\mathbf{H}_N-i\gamma{\bf e}\otimes {\bf e}^T$, with $\mathbf{H}_N$ taken from the Gaussian Orthogonal Ensemble[2,3]:
\begin{equation}\label{jpdGOE}
{\cal P}_N(z_1,\ldots,z_N)\propto \frac{e^{-\frac{N}{4}\left(\gamma^2+\sum_{j=1}^N(\Re z_j^2)\right)}}{\gamma^{\frac{N}{2}-1}}\prod_{j,k=1}^N\frac{1}{\sqrt{|z_j-z^*_k|}}\prod_{j<k}^N|z_j-z_k|^2\delta(\sum_{j=1}^N\Im z_j+\gamma)
\end{equation}

 Using it we follow essentially the same procedure of reduction to averaging over matrices of size $(N-1)\times (N-1)$  as in the GUE case (cf. also [4]) and come to the following relation:
\begin{equation}\label{Mom1GOE1}
{\cal O}^{(\beta=1)}_p(z)\propto \frac{1}{(\gamma_*Y)^{1/2}}\left(\frac{\gamma_*}{\gamma}\right)^{\frac{N}{2}-1}\,e^{-\frac{N}{4}(X^2-Y^2+\gamma^2)+\frac{N-1}{4}\gamma_*^2} {\cal M}^{(\beta=1)}_p(z)
\end{equation}
where
\begin{equation}\label{Mom2}
{\cal M}^{(\beta=1)}_p(z)=\left\langle\frac{\left[\det(\tilde{z}-\tilde{{\cal H}}_{eff})\det(\tilde{z}^*-\tilde{{\cal H}}^{\dagger}_{eff})\right]^{p+1}}{\left[\det(\tilde{z}-\tilde{{\cal H}}^{\dagger}_{eff})\det(\tilde{z}^*-\tilde{{\cal H}}_{eff})\right]^{p+\frac{1}{2}}}\right\rangle
\end{equation}
which, assuming for simplicity $X=0$ and the scaling $Y=y/N$,  asymptotically  in the limit $N\to \infty$ is equivalent to
\begin{equation}\label{Mom1GOE2}
{\cal O}^{(\beta=1)}_p(z)\propto \frac{e^{-\frac{y}{2}\left(\gamma+\frac{1}{\gamma}\right)}}{y^{1/2}}\lim_{N\to \infty} {\cal M}^{(\beta=1)}_p(z)
\end{equation}
where one replaces $\tilde{{\cal H}}_{eff}$ with the original ${\cal H}_{eff}=\mathbf{H}_N-i\gamma\,\tilde{{\bf w}}\otimes \tilde{{\bf w}}^{\dagger}$ without changing the result and may consider the  random real gaussian channel vector ${\bf \tilde{w}}$ rather than keeping it of unit length. The same steps as in the GUE case then give:
 \begin{equation}\label{Mom3GOE}
{\cal M}^{(\beta=1)}_p(z)=\left\langle |\det(z-\mathbf{H}_{N})|\frac{\left|1+i\gamma G(z)\right|^{2(p+1)}}{\left|1+i\gamma G(z^*)\right|^{2p+1}}\right\rangle
\end{equation}
 \begin{equation}\label{Mom4GOE}
=\int_{-\infty}^{\infty}du\int_0^{\infty}dv \frac{\left[(1-\gamma v)^2+\gamma^2 u^2\right]^{p+1}}{\left[(1+\gamma v)^2+\gamma^2 u^2\right]^{p+\frac{1}{2}}}{\cal P}_{z}(u,v)
\end{equation}
$$
{\cal P}^{(\beta=1)}_{z}(u,v)=\left\langle |\det(z-\mathbf{H}_{N})| \delta\left(u-\mbox{Re} G(z)\right)\,\delta\left(v-\mbox{Im} G(z)\right) \right\rangle
$$
After performing the Fourier-Laplace transform of the above function and averaging over random Gaussian real vectors $\tilde{\bf w}$
we reduce the problem to evaluating of the following function
\begin{equation}\label{Mom5GOEA}
 {\cal F}^{(\beta=1)}(s,p)=\left\langle\frac{|\det(\frac{i y}{N}-\mathbf{H}_{N})|^2}
 {\left[\det\left(-i\frac{\tau_{+}}{2N}-\mathbf{H}_{N}\right)\det\left(-i\frac{\tau_{-}}{2N}-\mathbf{H}_{N}\right)\right]^{\frac{1}{2}}}\right\rangle
\end{equation}
where  in the present case  $\frac{1}{2}\tau_{+}=s+\sqrt{s^2+y(y+2p)}>0$ and  $\frac{1}{2}\tau_{-}=s-\sqrt{s^2+4y(y+p)}<0$
and the averaging is to be performed over GOE matrices  $\mathbf{H}_{N}$ as $N\to \infty$.
Such a correlation function has been evaluated in [5], with the result being given by
\begin{equation}\label{Mom5GOEB}
 {\cal F}^{(\beta=1)}(s,p)\propto \frac{\sinh{2y}}{2y}\,\frac{|\tau_{+}-\tau_{-}|}{2}\,K_1\left(\frac{|\tau_{+}-\tau_{-}|}{4}\right)+\beta(p,q),
 \end{equation}
 where
 \[
 \beta(p,q)=\frac{2}{(2y)^3}\left[2y\cosh{2y}-\sinh{2y}\right]\left(y^2-\frac{1}{4}\tau_{+}\tau_{-}
 \right)K_0\left(\frac{|\tau_{+}-\tau_{-}|}{4}\right)
 \]
 Substituting here the values of $\tau_{\pm}$ one then arrives at
\begin{equation}\label{Mom5GOEC}
 {\cal F}^{(\beta=1)}(s,p)\propto 2\frac{\sinh{2y}}{2y}\,\sqrt{s^2+y(y+2p)}\,K_1\left(\sqrt{s^2+y(y+2p)}\right)+\beta(p,q),
 \end{equation}
 with
 \[
 \beta(p,q)=\frac{2}{(2y)^3}\left[2y\cosh{2y}-\sinh{2y}\right]\left(y+p\right)K_0\left(\sqrt{s^2+y(y+2p)}\right)
 \]
The Fourier-Laplace inversion is performed using the functions defined in (\ref{FL1}), namely
\begin{equation}
\phi_{-\frac{1}{2}}(s,p;y/2)=\frac{2}{\sqrt{y}}\,K_0(\sqrt{s^2+y^2+2py}), \quad \phi_{\frac{1}{2}}(s,p;y/2)=\frac{2}{y^{3/2}}\sqrt{s^2+y^2+2py}\,K_1(\sqrt{s^2+y^2+2py})
\end{equation}
which also can be used to prove the following identity:
\[
2p\,K_0(\sqrt{s^2+y^2+2py})=-\sqrt{y}\int_0^{\infty}e^{-pv}\,
\int_{-\infty}^{\infty}\frac{du}{\sqrt{2\pi}}e^{isu}\frac{e^{-\frac{y}{2}\frac{u^2+v^2+1}{v}}}{v^{\frac{3}{2}}}\left[\frac{3}{2v}+\frac{y}{2}
\left(2-\frac{u^2+v^2+1}{v^2}\right)\right]
\]
These identities when applied to the inversion of (\ref{Mom5GOEB}) yield
\begin{equation}\label{jpdGOE}
{\cal P}^{(\beta=1)}_{z}(u,v)\propto \left\{\left(\cosh{2y}-\frac{\sinh{2y}}{2y}\right)\left(-\frac{3}{y}+
\frac{u^2+v^2+1}{v^2}\right)+2\sinh{2y}\right\} \frac{\sqrt{y}}{4} \frac{1}{v^{\frac{5}{2}}}\, e^{-\frac{y}{2}\frac{u^2+v^2+1}{v}}
\end{equation}
and the final expression (17) for ${\cal P}^{(\beta=1)}_{y}(t)$ is straightforwardly obtained following the same manipulations as described for the GUE-based case.

Numerical simulations provide a reasonably good support for the final result,  Eq.(18) of the main text, already for relatively small matrices.
  \begin{figure}[h!]
    \includegraphics[width=0.5\textwidth,keepaspectratio=true]{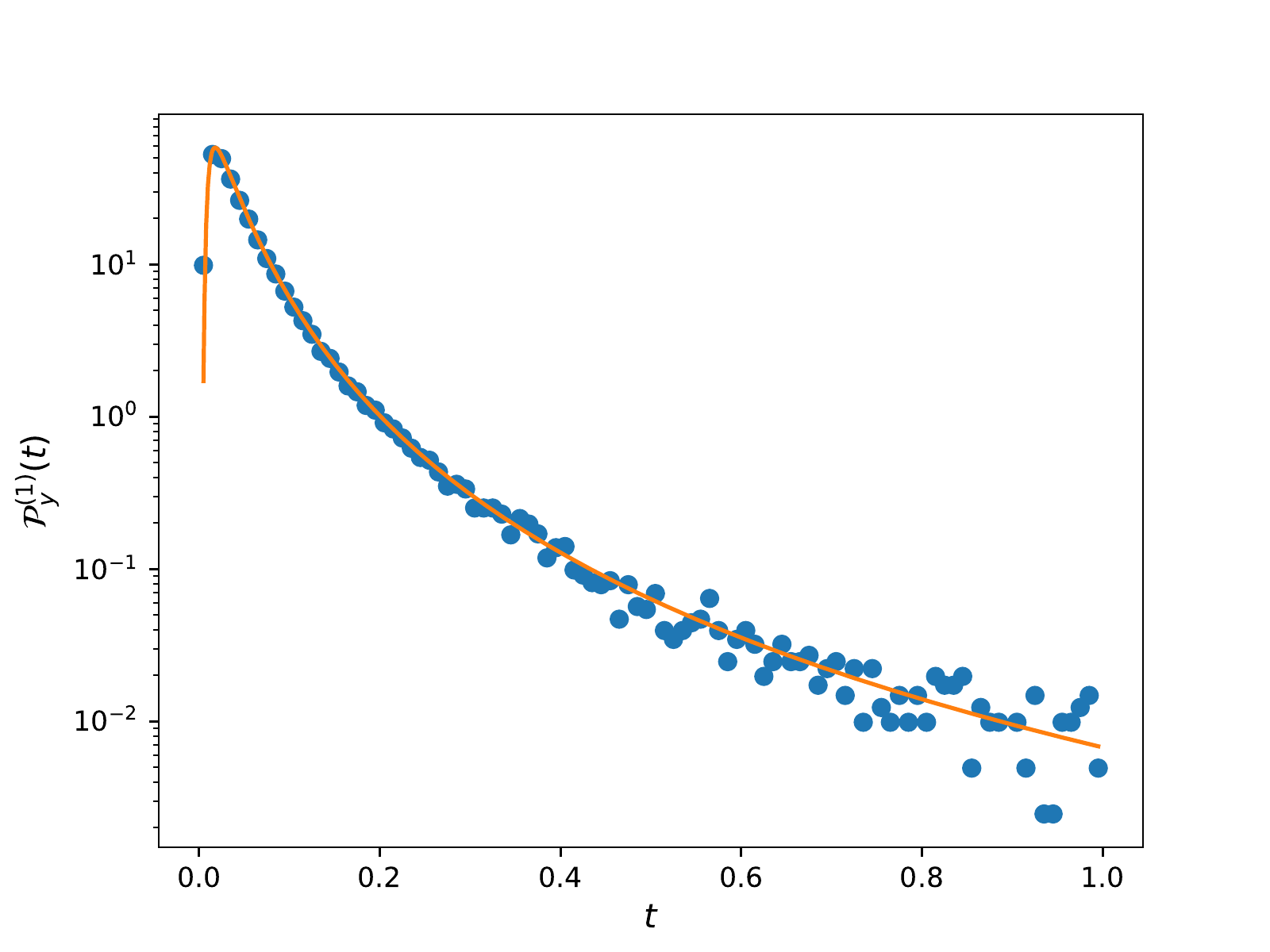}
        \label{fig:beta=1y=0.05}
        \caption{   Eq.18 compared with the empirical density of overlaps $O_{nn}$ for $n=10^{6}$ realizations of $32\times32$ matrices, with $y=0.05.$ and $\gamma=0.5$.}
\end{figure}

\subsection{Comparison of the formula Eq.(19) for the mean resonance density with the form in [6]}

We start with recalling  Eq.(4) from [6]:
\begin{equation}\rho^{(\beta=1)}(y)= \frac{1}{4\pi}\frac{d^2}{dy^2}
\int_{-1}^{1}(1-\lambda^2)e^{2\lambda y}(g-\lambda){\cal F}(\lambda,y)\,d\lambda
\end{equation}
where
\[
{\cal F}(\lambda,y)=\int_g^{\infty}dp_1\frac{e^{-y p_1}}{(\lambda-p_1)^2\sqrt{(p_1^2-1)(p_1-g)}}
\int_1^{g}dp_2\frac{e^{-y p_2}(p_1-p_2)}{(\lambda-p_2)^2\sqrt{(p_2^2-1)(g-p_2})}
\]
Below we consider only the perfect coupling case $g\to 1$. For this first change the variable: $p_2=1+(g-1)t, \, t\in[0,1]$,
implying $\frac{dp_2}{\sqrt{(p_2-1)(g-p_2)}}=\frac{dt}{\sqrt{t(1-t)}}$. This allows to evaluate the integral over $t$ in the limit
$g\to 1$, the result being equal to $\frac{\pi}{\sqrt{2}} e^{-y}\frac{(p_1-1)}{(\lambda-1)^2}$, yielding the formula:
\begin{equation}\label{SFT-perfect}
\rho^{(\beta=1)}_{g=1}(y)=\frac{1}{4\sqrt{2}}\int_1^{\infty}\frac{dp}{\sqrt{p+1}}e^{-y(p+1)}\int_1^{1}e^{2\lambda y}\frac{(1+\lambda)(2\lambda-p-1)^2}{(\lambda-p)^2}
\end{equation}
which should be compared to the $g=1$ limit of  Eq.(19)-(20) in the paper. Although we were not yet able to show analytically the equivalence of the two formulas, the numerical comparison is flawless:

 \begin{figure}[h!]
    \includegraphics[width=0.5\textwidth,keepaspectratio=true]{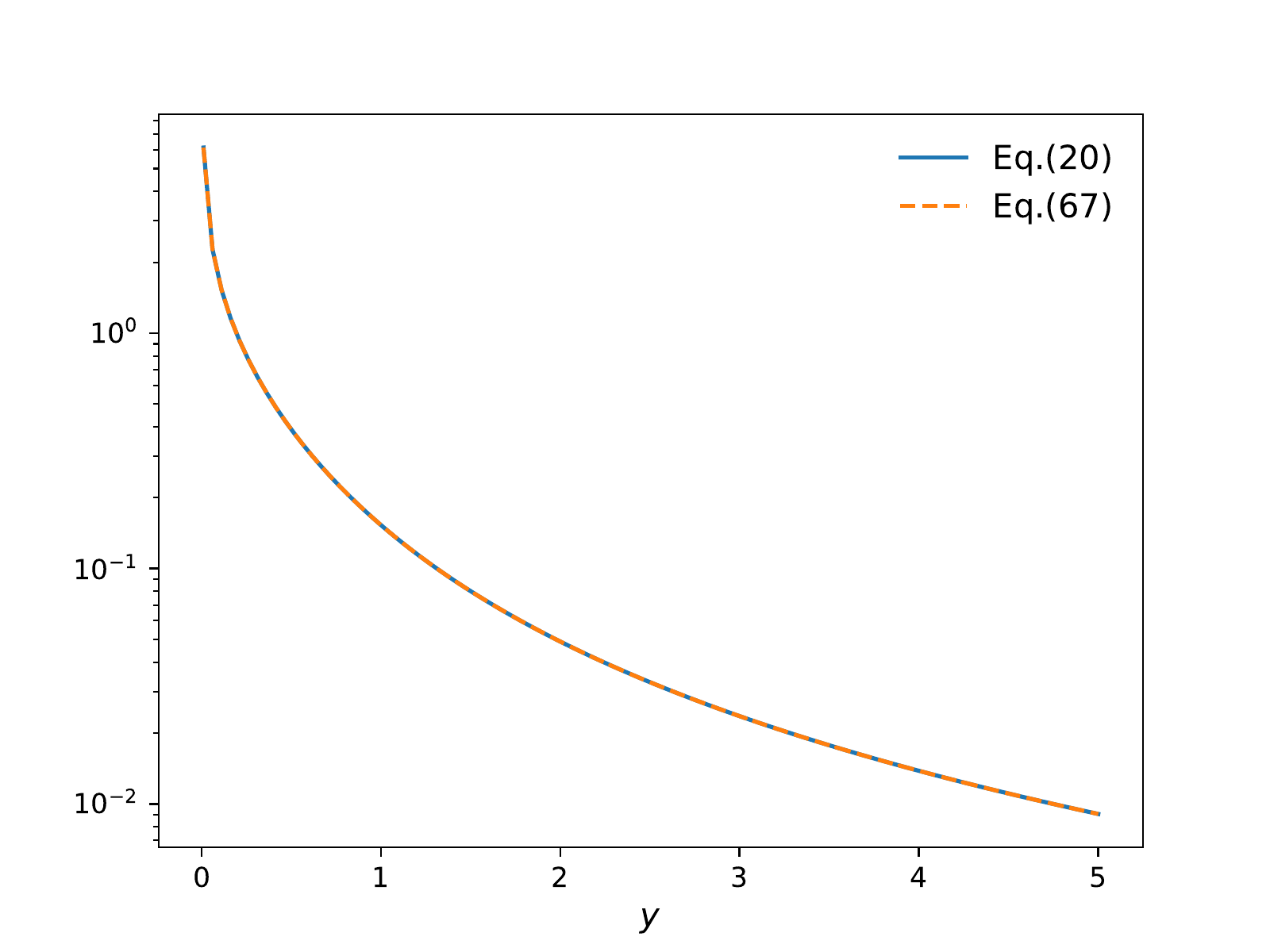}
        \caption{   Eq.(19) of the main text compared with Eq.(\ref{SFT-perfect})}
\end{figure}

\subsection*{Cited References}

\begin{enumerate}
\item A. Borodin A and E. Strahov. Averages of characteristic polynomials in random matrix theory.
{\it Commun. Pure Appl. Math.} {\bf 59} 161–-253 (2006)
\item N. Ullah. On a Generalized Distribution of the Poles of the Unitary Collision Matrix.
{\it J. Math.Phys.}{\bf 10} no. 11, 2099--2103 (1969)
\item R. Kozhan. Rank One Non-Hermitian Perturbations of Hermitian
$\beta$-Ensembles of Random Matrices. {\it J. Stat. Phys.} {\bf 168} 92--108 (2017)
\item Y. V. Fyodorov and D.V. Savin. Resonance width distribution in RMT:
Weak-coupling regime beyond Porter-Thomas. {\it EPL},\, {\bf 110},  40006, (2015)
\item  Yan V. Fyodorov, A. Nock.
On Random Matrix Averages Involving Half-Integer Powers of GOE Characteristic Polynomials.
{\it J. Stat. Phys.} {\bf 159}, {\bf  4}, 731-751 (2015)
\item H.-J. Sommers, Y. V.  Fyodorov, and M. Titov. S-matrix poles for chaotic quantum systems as eigenvalues of complex symmetric random
matrices: from isolated to overlapping resonances. {\it J. Phys. A: Math.
  Gen.},\,{\bf 32}, L77 (1999)
\end{enumerate}

\end{widetext}


\begin{thebibliography}{99}

\bibitem{Baranov_etal_CPA_2017}
D. G. Baranov, A. Krasnok, T. Shegai, A. Alu, and
Y. Chong. Coherent perfect absorbers: linear control
of light with light. Nature Reviews Materials {\bf 2}, 17064
(2017).

\bibitem{chong_coherent_2010}
 Y. D. Chong, Li Ge, Hui Cao and A.D. Stone. Coherent Perfect Absorbers: Time-Reversed Lasers.
 {\it Phys. Rev. Lett} {\bf 105}(5), 053901 (2010) [doi: 10.1103/PhysRevLett.105.053901]

\bibitem{Suwun_etal_2021} Suwun Suwunnarat, Yaqian Tang, Mattis Reisner, Fabrice Mortessagne, Ulrich Kuhl, and Tsampikos Kottos.
 Towards a Broad-Band Coherent Perfect Absorption in systems without
Scale-Invariance. {\it arxiv: 2103.03668}


\bibitem{pichler_random_2019}
  K. Pichler, M. K{\"u}hmayer, J. B{\"o}hm, A.  Brandst{\"o}tter, P. Ambichl, U. Kuhl,  and S. Rotter.
  Random Anti-Lasing through Coherent Perfect Absorption in a Disordered Medium. {\it Nature} {\bf 567} No. 7748, 351 (2019) [doi:10.1038/s41586-019-0971-3]

 \bibitem{chen_perfect_2020}
  L. Chen, T. Kottos, S.M. Anlage. Perfect Absorption in Complex Scattering Systems with or without Hidden Symmetries. 
  {\it Nat. Commun.} {\bf  11}: 5826 (2020)

 \bibitem{Delhoughne_etal_CPA} P del Hougne, KB Yeo, P Besnier, M Davy.
  On-demand coherent perfect absorption in complex scattering systems:
  time delay divergence and enhanced sensitivity to perturbations. {\it Arxiv:2010.06438}


\bibitem{li_random_2017}
  H. Li, S. Suwunnarat, R. Fleischmann, H. Schanz,  and T. Kottos. Random Matrix Theory Approach to Chaotic Coherent Perfect Absorbers.
  {\it Phys. Rev. Lett.} {\bf 118} No. 4, 044101 (2017)
  [doi:10.1103/PhysRevLett.118.044101]

\bibitem{fyodorov_distribution_2017}
 Y.V. Fyodorov, S. Suwunnarat, T. Kottos. Distribution of zeros of the S-matrix of chaotic cavities with localized losses and coherent perfect absorption: non-perturbative results. {\it  J. Phys. A: Math. Theor.} {\bf 50}, Issue 30,  30LT01 (2017)\,\, [doi:10.1088/1751-8121/aa793a]

 \bibitem{Baranov_etal_2017}  D. G. Baranov, A. Krasnok, and A. Alu. Coherent virtual absorption based on complex zero excitation for ideal light capturing. {\it Optica} {\bf 4}, Issue 12, 1457--1461 (2017)

 \bibitem{fyodorov_reflection_2019}
 Yan V. Fyodorov. Reflection Time Difference as a Probe of S-Matrix Zeroes in Chaotic Resonance Scattering.
   {\it Acta Physica Polonica A} {\bf 136} No.5, 785--789 (2019) [doi:10.12693/APhysPolA.136.785]

 \bibitem{OsmanFyodorov2020} Mohammed Osman, Yan V Fyodorov.
 Chaotic scattering with localized losses: $S-$matrix zeros and reflection time difference for systems with broken time-reversal invariance. {\it Phys. Rev. E} {\bf 102} (1), 012202 (2020)

\bibitem{Kang_Genak2021} Yuhao Kang and Azriel Z. Genack. Transmission zeros with topological symmetry in complex systems.
{\it Phys. Rev. B} {\bf 103}, L100201 (2021)


\bibitem{Doron1990} E. Doron, U. Smilansky and
A. Frenkel.  Experimental Demonstration of Chaotic Scattering of Microwaves. {\it Phys. Rev. Lett.} {\bf 65}, no. 25. 3072--3075 (1990)


\bibitem{Kuhletal2003} R.A. Mendez-Sanchez, U. Kuhl, M. Barth, C.V. Lewenkopf, and H.-J. St\"{o}ckmann. Distribution
of reflection coefficients in absorbing chaotic microwave cavities. {\it Phys. Rev. Lett.} {\bf 91}, 174102 (2003).

\bibitem{Hemmadi2005} S. Hemmady, X. Zheng, E. Ott, T.M. Antonsen, S.M. Anlage. Universal
impedance fluctuations in wave chaotic systems. {\it Phys. Rev. Lett.} {\bf 94}, 014102 (2005).


\bibitem{Hemmadi2006b} X. Zheng, S. Hemmady, T. M. Antonsen, Jr., S. M. Anlage, and E. Ott. Characterization of fluctuations of impedance and scattering matrices in wave chaotic scattering. {\it Phys. Rev. E} {\bf 73}, 046208 (2006).

\bibitem{microwgraphs2} M. Lawniczak, O. Hul, S. Bauch, P. Seba, and L. Sirko. Experimental and numerical investigation
of the reflection coefficient and the distributions of Wigner's reaction matrix for irregular graphs
with absorption. {\it Phys. Rev. E} {\bf 77}, 056210 (2008).

\bibitem{microwgraphs4} M. Lawniczak, L Sirko
Investigation of the diagonal elements of the Wigner’s reaction matrix for networks with violated time reversal invariance
{\it Scientific reports} {\bf 9} (1), 1-10 (2019)


  \bibitem{wigner55} E.P. Wigner. Lower limit for the energy derivative of the scattering phase
shift. {\it Phys. Rev.} {\bf 98} (1)  145–-147 (1955)

\bibitem{Lehmann95} N. Lehmann, D.V. Savin, V.V. Sokolov, H.-J. Sommers, Time delay correlations
in chaotic scattering: random matrix approach. {\it Physica D} {\bf 86}, 572–-585 (1995)

\bibitem{FSS96} Y.V. Fyodorov, H.-J. Sommers, Parametric correlations of scattering phase
shifts and fluctuations of delay times in few-channel chaotic scattering, {\it Phys.
Rev. Lett.} {\bf 76} (25), 4709 (1996)

\bibitem{FSS97}
 Y.V. Fyodorov, D.V. Savin, H.-J. Sommers, Parametric correlations of phase
shifts and statistics of time delays in quantum chaotic scattering: crossover
between unitary and orthogonal symmetries, {\it Phys. Rev. E} {\bf 55}
R4857–-R4860 (1997)

\bibitem{Brower99} P.W. Brouwer, K.M. Frahm, C.W. Beenakker. Distribution of the quantum
mechanical time-delay matrix for a chaotic cavity {\it Waves Random Media} {\bf 9} 91–-104 (1999)
\bibitem{SFS01}  D.V. Savin, Y.V. Fyodorov, H.-J. Sommers. Reducing nonideal to ideal coupling
in random matrix description of chaotic scattering: application to the timedelay problem. {\it Phys. Rev. E}
 {\bf 63} 035202 (2001)

\bibitem{OssFyo05}  A. Ossipov, Y.V. Fyodorov. Statistics of delay times in mesoscopic systems as a
manifestation of eigenfunction fluctuations. {\it Phys. Rev. B} {\bf 71}  125133 (2005)

\bibitem{Kott05} T. Kottos. Statistics of resonances and delay times in random media: beyond random matrix theory.
{\it J. Phys. A: Math. Gen.} {\bf 38}  10761–-10786 (2005)


\bibitem{MezSimm13} F. Mezzadri, N.J. Simm. $\tau$-Function theory of quantum chaotic transport with
$\beta = 1, 2, 4$. {\it Commun. Math. Phys.} {\bf 324},  465–-513 (2013)

\bibitem{TexMaj013} Christophe Texier and Satya N. Majumdar. Wigner Time-Delay Distribution in Chaotic Cavities and Freezing Transition.
{\it Phys. Rev. Lett.} {\bf 110}, 250602 (2013)

\bibitem{Novaes15} M. Novaes. Statistics of time delay and scattering correlation functions in
chaotic systems. I. Random matrix theory. {\it J. Math. Phys.} {\bf 56} 062110 (2015)

\bibitem{Cunden15} F.D. Cunden. Statistical distribution of the Wigner–Smith time-delay matrix
moments for chaotic cavities. {\it Phys. Rev. E} {\bf 91}, 060102 (2015)

\bibitem{Texier16}  C. Texier. Wigner time delay and related concepts: Application to transport in coherent conductors.
{\it Physica E} {\bf 82}, 16--33 (2016)

\bibitem{Grabsch18} A. Grabsch, D.V. Savin and C. Texier. Wigner–-Smith time-delay matrix in chaotic cavities with non-ideal contacts. {\it  J. Phys. A: Math. Theor.} {\bf 51}, 404001 (2018)

\bibitem{Grabsch20} A. Grabsch. Distribution of the Wigner–Smith time-delay matrix for chaotic cavities with
absorption and coupled Coulomb gases. {\it J. Phys. A: Math. Theor.} {\bf 53}, 025202 (2020).

\bibitem{Chen_Anlage_Fyodorov2021} L. Chen, S.M. Anlage and Y.V. Fyodorov.  Generalization of Wigner Time Delay to Sub-Unitary Scattering Systems. {\it arXiv:2101.08335}

\bibitem{SokZel89} V.V. Sokolov and V.G. Zelevinsky.
Dynamics and statistics of unstable quantum states.
{\it  Nucl. Phys. A}  {\bf 504}, Issue 3, 562--588(1989)

\bibitem{Fyodorov97}
Y.~V. Fyodorov and H.-J. Sommers. Statistics of resonance poles, phase shifts and time delays in quantum chaotic scattering: Random matrix approach for systems with broken time-reversal invariance. {\it J. Math. Phys.} \textbf{38}, Issue 4, 1918--1981 (1997) \,\,[doi:10.1063/1.531919]


\bibitem{Irotter09} I. Rotter. A non-Hermitian Hamilton operator and the physics of open quantum systems.
{\it J Phys A: Math. Theor} {\bf 42}, No. 15, 153001 (2009)

\bibitem{FSav11} Y.V. Fyodorov and D.V. Savin. Resonance scattering of waves in chaotic systems.
  in: {\it The Oxford Handbook of Random Matrix Theory}, edited by G.  Akemann et al. (Oxford University Press, 2011),
  pp. 703--722

\bibitem{kuhl13} U. Kuhl, O. Legrand, and F. Mortessagne. Microwave experiments using open chaotic cavities in the realm of the effective Hamiltonian formalism. {\it Fortschr. Phys.},\, {\bf 61},  414--419 (2013)

\bibitem{Schomerus2015} H. Schomerus. Random matrix approaches to open quantum systems.
in: {\it Stochastic Processes and Random Matrices: Lecture Notes of the Les Houches Summer School 2015},
edited by G. Schehr et al. (Oxford University Press, 2017), pp. 409--473

\bibitem{Fyodorov2016} Y.V. Fyodorov. Random Matrix Theory of resonances: An overview.
in: {\it 2016 URSI International Symposium on Electromagnetic Theory (EMTS), Espoo, Finland, IEEE}, 666--669
(2016)

\bibitem{Baez_etal_2008} G. Báez, M. Martínez-Mares, and R. A. Méndez-Sánchez. Absorption strength in absorbing chaotic cavities.
{\it Phys. Rev. E} {\bf 78}, 036208(2008)

 \bibitem{fyod96}
Y.~V. Fyodorov and H.-J. Sommers. Statistics of S-matrix poles in few-channel chaotic
scattering: Crossover from isolated to overlapping
resonances. {\it JETP Lett.},\,{\bf 63} 1026--1030 (1996).


\bibitem{fyodorov_systematic_1999}
Y. V. Fyodorov, B. A.  Khoruzhenko. Systematic Analytical Approach to Correlation Functions of Resonances
in Quantum Chaotic Scattering. {\it Phys. Rev.
  Lett.},\,{\bf 83}, 65 --68 (1999)

\bibitem{somm99} H.-J. Sommers, Y. V.  Fyodorov, and M. Titov. S-matrix poles for chaotic quantum systems as eigenvalues of complex symmetric random
matrices: from isolated to overlapping resonances. {\it J. Phys. A: Math.
  Gen.},\,{\bf 32}, L77 (1999)


 \bibitem{FyoSav15} Y. V. Fyodorov and D.V. Savin. Resonance width distribution in RMT:
Weak-coupling regime beyond Porter-Thomas. {\it EPL},\, {\bf 110},  40006, (2015)

 \bibitem{kuhl_resonance_2008}
  U. Kuhl, R. H{\"o}hmann, J Main, and H.-J. St{\"o}ckmann.
  Resonance Widths in Open Microwave Cavities Studied by Harmonic Inversion.
  {\it Phys. Rev. Lett.} {\bf 100}, 254201 (2008) [doi: 10.1103/PhysRevLett.100.254101]

\bibitem{CM1998} J. T. Chalker and B. Mehlig. Eigenvector Statistics in Non-Hermitian Random Matrix Ensembles.
{\it Phys. Rev. Lett.} \textbf{81}, 3367 (1998).

\bibitem{MC2000} B. Mehlig and J. T. Chalker. Statistical properties of eigenvectors in non-Hermitian Gaussian random matrix ensembles.
{\it J. Math. Phys.} {\bf 41}, 3233 (2000).

\bibitem{WaltersStarr2015} M. Walters and S. Starr. A note on mixed matrix moments for the complex Ginibre ensemble.
 {\it J. Math. Phys.} {\bf 56}, Issue 1, 013301 (2015)


\bibitem{Burda2015} Z. Burda, J. Grela, M.A. Nowak, W. Tarnowski, P. Warchoł.
Unveiling the significance of eigenvectors in diffusing non-Hermitian matrices by identifying the underlying Burgers dynamics
{\it Nucl. Physics B} {\bf 897}, 421--447 (2015)

\bibitem{Belinchi2017} S. Belinschi, M. A. Nowak, R. Speicher and W. Tarnowski.
Squared eigenvalue condition numbers and eigenvector correlations from the single ring theorem.
{\it J Phys. A: Math. Theor.}, {\bf 50} 105204 (2017)


\bibitem{BD2018}
P. Bourgade, G. Dubach. The distribution of overlaps between eigenvectors of Ginibre matrices.
{\it Probab. Theory Relat. Fields} {\bf 177} 397–-464 (2020)

\bibitem{Fyod2018} Y. V. Fyodorov. On statistics of bi-orthogonal eigenvectors in real and complex Ginibre ensembles: combining partial Schur decomposition with supersymmetry.  {\it Commun. Math. Phys.} {\bf 363}, 579–-603 (2018).


\bibitem{GW2018}
J. Grela, P. Warchoł. Full Dysonian dynamics of the complex Ginibre ensemble.
{\it J. Phys. A: Math. Theor.} \textbf{51}, 42 (2018).

\bibitem{NW2018} M.A. Nowak, W. Tarnowski. Probing non-orthogonality of eigenvectors in non-Hermitian matrix models: diagrammatic approach. {\it J. High Energ. Phys.} {\bf 2018}, 152 (2018).

 \bibitem{Zeit2018} F. Benaych-Georges, O. Zeitouni. Eigenvectors of non normal random matrices
{\it Electron. Commun. Probab.} {\bf 23} 1--12 (2018).

\bibitem{Dubach2019} G Dubach. On eigenvector statistics in the spherical and truncated unitary ensembles
 {\it arXiv:1908.06713}

\bibitem{Metz2019}
F. L. Metz, I. Neri and T. Rogers.
Spectral theory of sparse non-Hermitian random matrices
 {\it J. Phys. A: Math. Theor.} {\bf 52} 434003 (2019)

\bibitem{GNetal2020}
E. Gudowska-Nowak, J. Ochab, D. Chialvo, M. A. Nowak and W. Tarnowski. From synaptic interactions to collective dynamics in random neuronal networks models: critical role of eigenvectors and transient behavior. {\it Neural Computation} {\bf 32} (2): 395–423 (2020)

\bibitem{Akemann2020a}
G. Akemann, R. Tribe, A. Tsareas, O. Zaboronski.
On the determinantal structure of conditional overlaps for the complex Ginibre ensemble
{\it Random Matrices: Theory and Applications} {\bf 09}, No. 04, 2050015 (2020)

\bibitem{Akemann2020b}
G. Akemann, Y.-P. F\"{o}rster and M. Kieburg.
Universal eigenvector correlations in quaternionic Ginibre ensembles
{\it J. Phys. A: Math. Theor.} {\bf  53}, Number 14, 145201 (2020)

\bibitem{FyoTarn2021}
Y. V. Fyodorov and  W. Tarnowski.
Condition Numbers for Real Eigenvalues in the Real Elliptic Gaussian Ensemble
{\it Annales Henri Poincaré} {\bf 22}, 309–-330(2021)

\bibitem{SavSok97}  D.V. Savin and V.V. Sokolov. Quantum versus classical decay laws in open chaotic systems.
{\it Physical Review E} {\bf 56} (5), R4911 (1997)


 \bibitem{jani99} R. A. Janik, W. Noerenberg, M. A. Nowak, G. Papp, I. Zahed. Correlations of Eigenvectors for Non-Hermitian Random-Matrix Models. ,\,{\it Phys. Rev. E} {\bf 60}, 2699 (1999)

\bibitem{scho00} H. Schomerus, K.M. Frahm, M. Patra, C.W.J. Beenakker. Quantum limit of the laser line width in chaotic cavities and statistics of residues of scattering matrix poles. {\it Physica A},\, {\bf 278}, 469--496 (2000)

\bibitem{MehlSant2001} B. Mehlig and M. Santer. Universal eigenvector statistics in a quantum scattering ensemble.
{\it Phys. Rev. E} {\bf 63}, 020105(R) (2001)

\bibitem{FyoMehl2002} Y.V. Fyodorov and B. Mehlig. Statistics of resonances and nonorthogonal eigenfunctions in a model for single-channel chaotic scattering. {\it Phys. Rev. E}, {\bf 66}, 045202(R) (2002)

\bibitem{fyodorov_random_2003}
 Yan V. Fyodorov and H.-J. Sommers.  Random Matrices Close to Hermitian or Unitary: Overview of Methods and Results.
  {\it J. Phys. A: Math. Gen.} {\bf 36}, no.12, 3303--3347 (2003)\, \, [doi: 10.1088/0305-4470/36/12/326]

\bibitem{pert} C. Poli, D. V. Savin, O. Legrand, and F. Mortessagne. Statistics of resonance states in open chaotic systems: A perturbative approach
{\it Phys. Rev. E} {\bf 80}, 046203 (2009)

 \bibitem{FyoSav2012} Y.V. Fyodorov and D.V. Savin. Statistics of resonance width shifts as a signature of eigenfunction nonorthogonality.
 {\it Phys. Rev. Lett.} {\bf  108}, 184101 (2012)


\bibitem{gros14} J.-B. Gros, U. Kuhl, O. Legrand, F. Mortessagne, E. Richalot, and D. V. Savin. Experimental width shift distribution: a test of nonorthogonality for local and global perturbations. {\it Phys. Rev. Lett.},\, {\bf 113} 224101 (2014)


\bibitem{DavyGenack18} M. Davy and A. Z. Genack. Selectively exciting quasi-normal modes in open disordered systems.
{\it Nature Commun.} {\bf 9}, Article number: 4714 (2018)

\bibitem{DavyGenack19} M. Davy and A. Z. Genack. Probing nonorthogonality of eigenfunctions and its impact on transport through open systems. {\it Physical Review Research}{\bf 1}, 033026 (2019)

\bibitem{SM} see Supplementary Materials section.

\bibitem{VWZ85}
J. J.~M. Verbaarschot, H.A. Weidenm{\"{u}}ller and M.R. Zirnbauer. Grassmann integration in stochastic quantum physics: the case of compound-nucleus scattering.  {\it Phys. Rep.},\,{\bf 129}, 367--438 (1985)

\bibitem{Fyo2000} Y.V. Fyodorov. Spectra of random matrices close to unitary and scattering theory for
discrete-time systems. in: {\it Disordered and complex systems}, AIP conference proceedings
{\bf 553}, pages 191–-196. (Amer. Inst. Phys., Melville, NY, 2001)

\bibitem{FyoSom2000} YV Fyodorov, HJ Sommers. Spectra of random contractions and scattering theory for discrete-time systems
{\it JETP Letters} {\bf 72} (8), 422--426 (2000)

\bibitem{ZS2000} K. Zyczkowski and H.-J. Sommers. Truncations of random unitary matrices. {\it J. Phys. A}
{\bf 33}, 2045–-2057 (2000)

\bibitem{StSe98}  H.-J. St\"{o}ckmann and P. Seba. The joint energy distribution function for the
Hamiltonian $H_0-iWW^{\dagger}$ for the one-channel case. {\it J. Phys. A: Math. Gen.} {\bf 31}, 3439-3448 (1998)

\bibitem{Kozhan07} R. Kozhan. Rank One Non-Hermitian Perturbations of Hermitian
$\beta$-Ensembles of Random Matrices. {\it J. Stat. Phys.} {\bf 168} 92--108 (2017)


\bibitem{MF1994} A.D. Mirlin and Y.V. Fyodorov. Statistical properties of one-point Green functions in disordered systems and critical behaviour near the Anderson transition.  {\it J. Phys. I France} {\bf 4}, 655--673 (1994)

\bibitem{fyodsav04} Y.V. Fyodorov, D.V. Savin. Statistics of impedance, local density of states, and reflection in quantum chaotic systems with absorption
{\it JETP Letters} {\bf 80} (12), 725--729 (2004)

\bibitem{savfyodsom05} D.V. Savin, H.J. Sommers, Y.V. Fyodorov.
Universal statistics of the local Green’s function in wave chaotic systems with absorption
{\it JETP Letters} {\bf 82} (8), 544--548 (2005)

\bibitem{AS1995} A.V. Andreev, B.D. Simons. Correlators of spectral determinants in quantum chaos. {\it Phys. Rev.
Lett.} {\bf 75}, 2304–-2307 (1995)


 \bibitem{FyoStra} Y.V. Fyodorov, E. Strahov. An exact formula for general spectral correlation function of
random Hermitian matrices. {\it J. Phys. A} {\bf 36}, 3203--3213
(2003)

\bibitem{StraFyo} E. Strahov, Y.V. Fyodorov. Universal results for correlations of characteristic polynomials:
Riemann–Hilbert approach. {\it Commun. Math. Phys.}
{\bf 241}, 343 (2003).

\bibitem{BorStra} A. Borodin A and E. Strahov. Averages of characteristic polynomials in random matrix theory.
{\it Commun. Pure Appl. Math.} {\bf 59} 161–-253 (2006)

\bibitem{FyoNock2015} Yan V. Fyodorov, A. Nock.
On Random Matrix Averages Involving Half-Integer Powers of GOE Characteristic Polynomials.
{\it J. Stat. Phys.} {\bf 159}, {\bf  4}, 731-751 (2015)

\bibitem{footnote} The case of symplectic symmetry is to be considered
in the paper: Y.V. Fyodorov and R. Tublin. A few results and conjectures about rank-one non-Hermitian deformations of $\beta$-Hermite Ensembles, {\it under preparation}.


\end{thebibliography}
\end{document}